\documentclass[apj,tighten,iop]{emulateapj}		

\usepackage{apjfonts}
\usepackage[]{natbib}
\usepackage{graphicx}	
\usepackage{amsmath}
\usepackage[T1]{fontenc}		
\usepackage{color}			
\usepackage[normalem]{ulem}

\newcounter{species}
\def\ion#1#2{\setcounter{species}{#2}#1$\;${\sc\roman{species}}\relax}

\def\fion#1#2{[{\setcounter{species}{#2}#1$\;${\sc\roman{species}}\relax}]}
\def\flion#1#2#3{[{\setcounter{species}{#2}#1$\;${\sc\roman{species}}]$\;\lambda${#3}}\relax}
\def\fllion#1#2#3{[{\setcounter{species}{#2}#1$\;${\sc\roman{species}}]$\;\lambda\lambda${#3}}\relax}

\newcommand{\kms}{km~s$^{-1}$}

\newcommand{\HeIIw}{He\,{\sc ii}~$\lambda$4686}

\newcommand{\OIIIw}{[O$\,\textsc{iii}]$~$\lambda$5007}
\newcommand{\OIIIdblt}{[O$\,\textsc{iii}]$~$\lambda\lambda$4959, 5007}

\newcommand{\HeII}{He\,{\sc ii}}

\newcommand{\Hb}{{H}$\beta$}
\newcommand{\FeII}{Fe\,{\sc ii}}

\newcommand{\OIII}{[O\,{\sc iii}]}

\def\lsim{\lower0.3em\hbox{$\,\buildrel <\over\sim\,$}}
\def\gsim{\lower0.3em\hbox{$\,\buildrel >\over\sim\,$}}

\newcommand{\pone}{Paper~\textsc{I}}
\newcommand{\ptwo}{Paper~\textsc{II}}

\def\kms{\,km~s$^{-1}$}      

\def\lesssim{\mathrel{\hbox{\rlap{\hbox{%
 \lower4pt\hbox{$\sim$}}}\hbox{$<$}}}}
\def\gtrsim{\mathrel{\hbox{\rlap{\hbox{%
 \lower4pt\hbox{$\sim$}}}\hbox{$>$}}}}

\def\arcsec{\hbox{$^{\prime\prime}$}}

\def\farcs{\hbox{$.\!\!^{\prime\prime}$}}

\shorttitle{Flux Variability of Supermassive Black Hole Binary Candidates }
\shortauthors{Runnoe et al.}

\begin{document}		
\title{A Large Systematic Search for Close Supermassive Binary and Rapidly Recoiling Black Holes - II. Continued Spectroscopic Monitoring and Optical Flux Variability}

\author{
Jessie C. Runnoe\altaffilmark{1,2}, Michael Eracleous\altaffilmark{1,3,4}, Gavin Mathes\altaffilmark{1,5}, Alison Pennell\altaffilmark{1}, Todd Boroson\altaffilmark{6}, Steinn Sigur{\dh}sson\altaffilmark{1}, Tamara Bogdanovi\'c\altaffilmark{3}, Jules P. Halpern\altaffilmark{7}, and Jia Liu\altaffilmark{7}}
\altaffiltext{1}{Department of Astronomy and Astrophysics and Institute for Gravitation and the Cosmos,
The Pennsylvania State University, 525 Davey Lab, University Park, PA 16803}

\altaffiltext{2}{Visiting astronomer, Kitt Peak National Observatory, National Optical Astronomy Observatory, which is operated by the Association of Universities for Research in Astronomy (AURA) under a cooperative agreement with the National Science Foundation.}

\altaffiltext{3}{Center for Relativistic Astrophysics, School of Physics, Georgia Institute of Technology, Atlanta, GA 30332}
\altaffiltext{4}{Department of Astronomy, University of Washington, Box 351580, Seattle, WA 98195}

\altaffiltext{5}{Department of Astronomy, New Mexico State University, Las Cruces, NM 88003}

\altaffiltext{6}{Las Cumbres Observatory Global Telescope Network, Goleta, CA 93117}

\altaffiltext{7}{Columbia Astrophysics Laboratory, Columbia University, 550 West 120th Street, New York, NY 10027-6601}

\begin{abstract}
We present new spectroscopic observations that are part of our continuing monitoring campaign of 88 quasars at $z<0.7$ whose broad \Hb\ lines are offset from their systemic redshifts by a few thousand km~s$^{-1}$. These quasars have been considered candidates for hosting supermassive black hole binaries (SBHBs) by analogy with single-lined spectroscopic binary stars. We present the data and describe our improved analysis techniques, which include an extensive evaluation of uncertainties. We also present a variety of measurements from the spectra that are of general interest and will be useful in later stages of our analysis. Additionally, we take this opportunity to study the variability of the optical continuum and integrated flux of the broad \Hb\ line. We compare the variability properties of the SBHB candidates to those of a sample of typical quasars with similar redshifts and luminosities observed multiple times during the Sloan Digital Sky Survey. We find that the variability properties of the two samples are similar (variability amplitudes of 10--30\% on time scales of approximately 1--7~years) and that their structure functions can be described by a common model with parameters characteristic of typical quasars. These results suggest that the broad-line regions of SBHB candidates have a similar extent as those of typical quasars. We discuss the implications of this result for the SBHB scenario and ensuing constraints on the orbital parameters.
\end{abstract}

\section{Introduction}
The empirical relationship between masses of supermassive black holes (BHs) and the stellar velocity dispersion or integrated stellar luminosity of their host galaxies \citep{kormendy93,magorrian98,ferrarese00a,gebhardt00,gultekin09} has been taken as an indication of an evolutionary connection between the black hole and host galaxy. The discovery of this relationship has motivated the development of a new generation of merger-driven galaxy evolution models in which coordinated black hole accretion and star formation in the host galaxy lead to the observed relation between their masses \citep[e.g.,][and references therein]{volonteri03,dimatteo05,hopkins06,volonteri15}. The theoretical models are bolstered by observations of dual active galactic nuclei (AGNs), with kiloparsec-scale separations, in interacting and merging galaxies. These can be detected directly via imaging \citep[e.g.,][]{junkkarinen01,komossa03,ballo04,comerford09b,fabbiano11,koss11,koss12} and indirectly via their double-peaked narrow emission lines \citep{comerford09a,wang09,smith10,liu10b,liu10a,shen11b,fu11a,fu12,barrows12,comerford12,comerford13}.

The formation of supermassive black holes binaries (SBHBs) during the late stages of the mergers is one outcome of hierarchical galaxy evolution. In the evolutionary scenario of SBHBs, first described in detail by \citet{begelman80}, the orbital decay of {\it close, bound} binaries slows down dramatically or stalls at separations of order 1~pc because of ``loss cone'' depletion (the loss cone is the region in stellar phase space containing stars that can scatter off the SBHB and drain its orbital angular momentum). This slow-down has been termed the ``last parsec problem.'' However, modern calculations have shown that the SBHB orbital decay may proceed unimpeded because the loss cone can be replenished by 3-body relaxation \citep[e.g.,][]{yu05,milos03}, the influence of non-spherical potentials \citep[e.g.,][]{yu02a,merritt04a,berczik06,khan13,vasiliev13,vasiliev14}, or interactions between the binary and a reservoir of gas \citep[e.g.,][]{armitage02,escala04,dotti07,dotti09b,cuadra09,lodato09,farris15b}.

Observational evidence for close, bound SBHBs has proven quite elusive. The two noteworthy candidates are CSO~0402$+$379, with a projected separation of approximately 7~pc \citep[imaged by radio interferometry][]{maness04,rodriguez06,rodriguez09} and OJ~287, where outbursts in the light curve occurring on a $\sim$12~year period have been explained in this context \citep[e.g.,][and references therein]{valtonen12}.  The identification of such systems is particularly difficult because these small separations are not resolved at cosmological distances. Yet, finding SBHBs is important because it would validate the galaxy evolution scenarios mentioned above, because SBHBs have been invoked to explain a variety of other astronomical observations \citep[see summary in][hereafter \pone]{eracleous12}, and because they are the progenitors of low-frequency gravitational wave sources potentially detectable by Pulsar Timing Arrays (e.g., \citealt{arzoumanian14s}, see \citealt{sesana15} for a review) and future space-based instruments \citep[see][for a review]{amaro-seoane12}.

Recent observational searches for sub-parsec SBHBs, motivated by the work of \citet{komossa08} and \citet{boroson10}, have turned up many new candidates.  An in-depth review of the field can be found in \citet{bogdanovic15} and a discussion of an array of techniques to identify SBHBs, primarily via their emission-lines, is given in \citet{popovic12}.  The widespread approach has been to make an analogy with single-line spectroscopic binary stars \citep[see e.g.][for a review of this concept]{gaskell96a} and search for indications of orbital motion in the periodic velocity shifts of the broad emission lines relative to the narrow lines. In this spirit, a number of groups have searched for quasars from the Sloan Digital Sky Survey (SDSS) whose broad \Hb\ or \ion{Mg}{2} emission lines are displaced from their systemic redshifts by several hundred to a few thousand \kms\ and/or show substantial radial velocity variations about their systemic redshifts \citep[\pone;][]{tsalmantza11, decarli13, shen13a, liu14, ju13}. In the working hypothesis adopted for these searches, the SBHB sits in a circumbinary disk and the secondary, which can more easily access gas on the inner edge of this disk \citep[see, for example,][]{artymowicz96,hayasaki07}, accretes at a much higher rate than the primary and alone has a broad-line region (BLR). In the picture emerging from the observations, SBHB candidates that have been extensively tested and are still widely held to be viable are rare, but initial studies of pairs of spectra, taken years apart, suggest that in some cases observed velocity shifts of the broad emission lines are consistent with orbital motion \citep[e.g., \pone][]{shen13a,ju13,liu14}.  It is important to note that the interpretation of these measurements is complicated by many significant caveats, as discussed in \pone, because the broad-line shifts are not unique signatures of orbital motion.

Obtaining more robust constraints on the nature of the SBHB candidates via the radial-velocity technique can be accomplished by long-term spectroscopic monitoring. If the orbital periods are of order a century, as noted by a number of authors \citep[e.g.,][\pone]{yu02a,loeb10,decarli13}, we may be able to detect small but systematic radial velocity variations over the course of 1--2 decades. If the orbital periods happen to be of order a few decades, we may be able to monitor a complete orbital cycle in our lifetimes. Thus, we have been carrying out a spectroscopic monitoring program of the sample from \pone\ and refining our methodology for analyzing the data.  In this paper, our primary goal is to present the spectra we have obtained since \pone\ and a refined analysis aimed at (a) verifying some of the results presented in \pone\ and (b) quantifying the uncertainties in our measurements of spectroscopic properties. We also take advantage of this opportunity to pursue a secondary goal: explore the variability of the continuum and broad \Hb\ line fluxes of our SBHB candidates and compare them to those of a sample of typical quasars of similar luminosity and redshift. In a forthcoming paper we concentrate on the study of long-term radial velocity variations, where we will adopt lessons learned here about broad line variability.

The continuum emission in quasars is known to vary, the amplitude of fluctuations is typically on the order of 10--30\%, increasing towards longer time scales and eventually leveling off at periods longer than a few years \citep[as inferred from long-term light curves and structure functions; e.g.,][and references therein]{hawkins93,giveon99,sesar06,macleod12}.  However, the mechanism producing the offset emission lines in the absence of a SBHB may leave its imprint on the flux variability as well. Recoiling black holes, where the supermassive black hole and BLR are kicked out of the center of the host galaxy following a merger, can also produce offset broad lines and would presumably display variability of the continuum and broad emission line flux that is consistent with a typical quasar.  Although simulations suggest that the accretion history of recoiling black holes will differ significantly from ordinary AGN over the course of their 10--100~Myr lifetimes \citep{blecha11,guedes11}, on time scales of decades or less the variability properties should be normal.  It is not completely clear what might be expected of flux variability in a SBHB.  Simulations of massive black hole binaries with gaseous disks generally find variable accretion rates \citep[e.g.,][]{hayasaki07,cuadra09,noble12,roedig12,shi12,bode12,dorazio13,farris14}, and calculations suggest that the luminosity and accretion rate can remain proportional \citep{farris15a}, but beyond this the results diverge.  The time scale for variability can range from 0.5 to 10 times the binary period \citep[e.g.,][]{dorazio13,farris14}. Observational searches have taken advantage of the possibility that continuum variability may be synchronized with orbital motion of the binary and some SBHB candidates have been claimed based on their light curve \citep[e.g.,][]{graham15,liu15}. In these cases the orbital periods are very short, amounting to a few years or less. Taken at face value, such systems would represent the last stage of the evolution of the SBHB, just before the merger (in comparison, the candidates found via offset broad emission lines have orbital periods of order a few hundred years and, if their orbits were to decay by gravitational radiation alone, tit would take them  $\sim 10^{10}\;$years to reach periods of a few years). However, the ubiquity of red noise in quasar power spectra introduces substantial uncertainty into this approach \citep{macleod10}.  With these considerations in mind, studying the flux variability properties of objects with offset broad emission lines can be valuable and informative, whether the line offsets are indicative of binaries, recoiling black holes, or an extreme population of individual quasars.  
		
In Section~\ref{sec:data} we summarize the sample selection and describe the acquisition of new data and the construction of a comparison sample of typical quasars.  In Section~\ref{sec:analysis} we assess the ensemble variability of the binary candidates and comparison sample, both by considering the changes in the integrated \Hb\ and optical continuum fluxes and by comparing structure functions for the two samples.  We discuss our results in the context of previous work and the binary hypothesis in Section~\ref{sec:implications}, and also consider the constraints implied by the observed variability on the properties of the binaries.  We summarize this investigation in Section~\ref{sec:summary}.  Additionally, Appendix~\ref{app:measurements} provides spectral measurements that have been made for the sample of SBHB candidates but are not utilized in this work.  Throughout this work, we adopt a cosmology of $H_0 = 73\;{\rm km\; s^{-1}\;Mpc^{-1}}$, $\Omega_{\Lambda} = 0.73$, and $\Omega_{m} = 0.27$.

\clearpage
\section{Data}
\label{sec:data}

\subsection{The Sample of Supermassive Black Hole Binary Candidates and New Spectra}
\label{sec:bbh_sample}
\begin{figure}[t]
\epsscale{1.0} \plotone{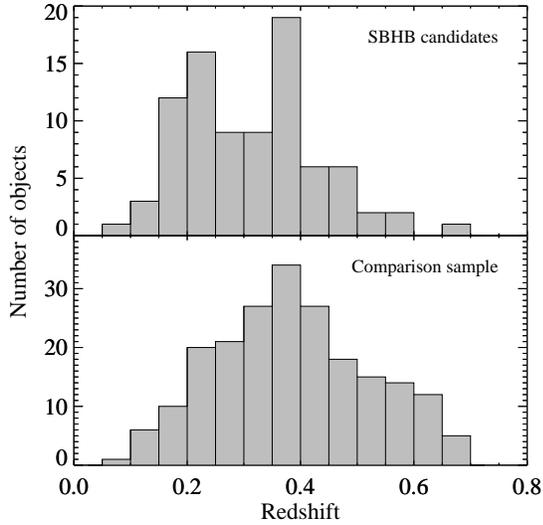} 
\caption{Distribution of redshifts for the SBHB candidates and the comparison sample of typical quasars from the SDSS. The two distributions are clearly not identical but they span approximately the same range and have approximately the same median.} \label{fig:zhist}
\end{figure}

The selection process and spectroscopic properties of the sample of SBHB candidates are described in detail in \pone. In summary, the sample was selected via spectral principal component analysis \citep[following the technique described in][]{boroson10} to identify a population of objects from the SDSS whose broad \Hb\ profiles do not resemble those of typical quasars.  We then vetted the initial, automatically-selected sample by visual inspection, discarding, for example, clearly double-peaked profiles \citep[which are likely produced by an accretion disk around a single supermassive BH, e.g.,][]{eracleous97, eracleous03, eracleous09}.  The final sample contains 88 objects having a median redshift of 0.32, with a full range of $0.077<z< 0.713$, as shown in the top panel of Figure~\ref{fig:zhist}.  Absolute $V$-band magnitudes, after Galactic extinction corrections but not K corrections, range from $-21.25$ to $-26.24$, with a median value for the sample of $-23.04$\footnote{We use ``quasar'' and AGN interchangeably in this work, even though some objects in this sample do not meet the strict absolute magnitude criterion for quasars.}.    

\begin{figure}[t]
\epsscale{1.3} \plotone{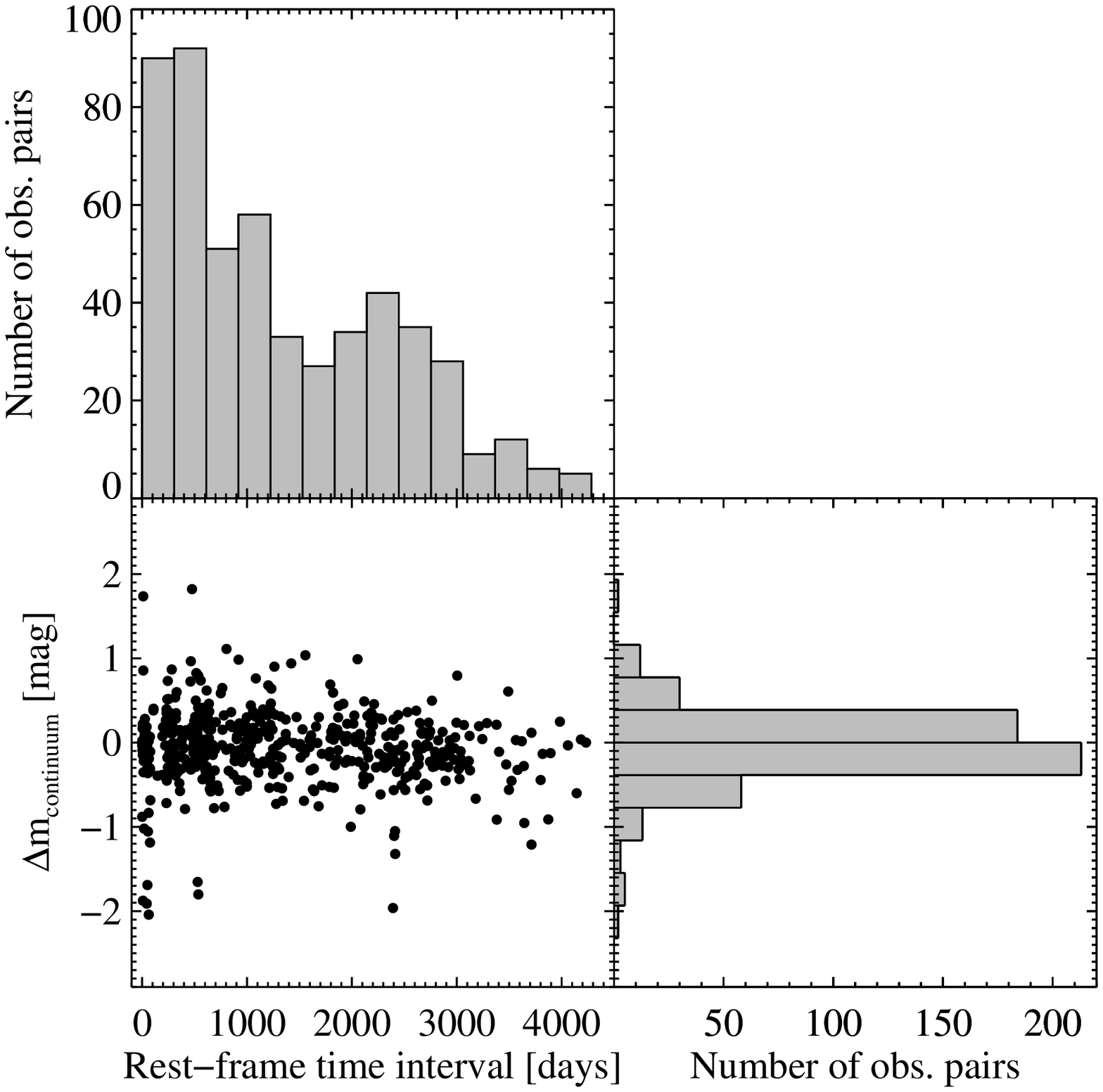} 
\caption{For every unique pair of observations for the SBHB candidates, the distribution of rest-frame time intervals (top), $\Delta m=-2.5\textrm{log}(f_1/f_2)$ measured from continuum flux density ratios at 5100~\AA\ (bottom right), and $\Delta m$ at 5100~\AA\ as a function of rest-frame time interval between observations (bottom left).  The $\Delta m$ distribution has a median of $-0.05$ and a standard deviation of 0.43.} \label{fig:dfdt_c}
\vfill
\epsscale{1.3} \plotone{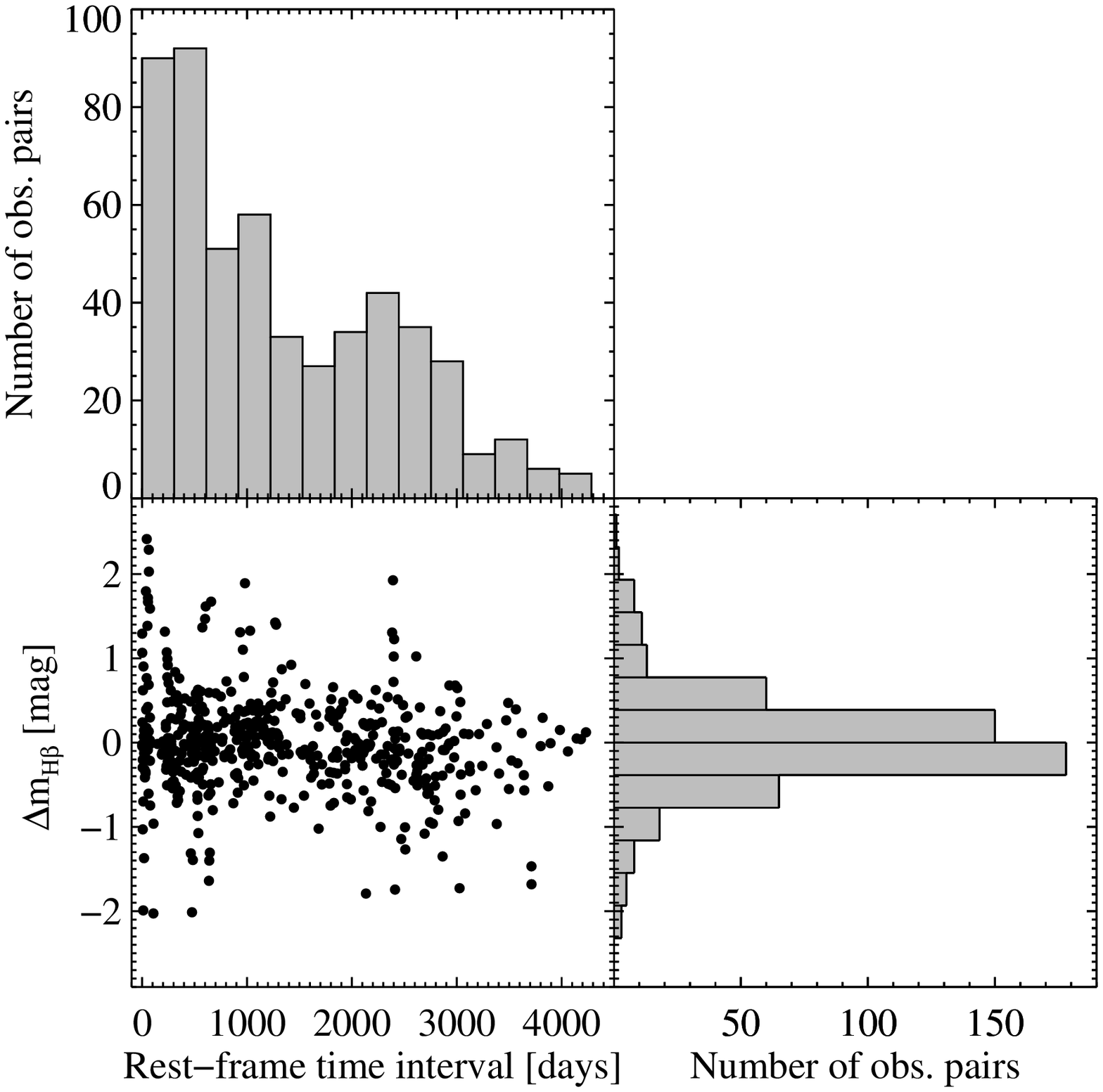} 
\caption{For every unique pair of observations for the SBHB candidates, the distribution of rest-frame time intervals (top), $\Delta m=-2.5\textrm{log}(f_1/f_2)$ measured from integrated \Hb\ flux ratios (bottom right), and the integrated \Hb\ $\Delta m$ as a function of rest-frame time interval between observations (bottom left).  The $\Delta m$ distribution has a median of $-0.03$ and a standard deviation of 0.59.} \label{fig:dfdt}
\end{figure}

\def\th{\tablenotemark{h}}
\begin{deluxetable*}{lcccccccccc}
\tablewidth{0in}
\tabletypesize{\scriptsize}
\tablecolumns{11}
\tablecaption{Log of Spectroscopic Observations [Abridged]\label{tab:obslog}}
\tablehead{
\colhead{} &
\colhead{} &
\colhead{} &
\colhead{} &
\colhead{} &
\colhead{} &
\colhead{} &
\colhead{} &
\colhead{Exposure} &
\colhead{} &
\colhead{Rest-Frame} \\
\colhead{Object Name} &
\colhead{} &
\colhead{$m_{\rm V}$\tablenotemark{b}} &
\colhead{$A_{\rm V}$\tablenotemark{c}} &
\colhead{$M_{\rm V}$\tablenotemark{d}} &
\colhead{No.} &
\colhead{Observation} &
\colhead{Instr.} &
\colhead{Time} &
\colhead{} &
\colhead{Wavelength} \\
\colhead{SDSS J} &
\colhead{$z$\tablenotemark{a}} &
\colhead{(mag)} &
\colhead{(mag)} &
\colhead{(mag)} &
\colhead{obs.\tablenotemark{e}} &
\colhead{Date (UT)} &
\colhead{Code\tablenotemark{f}} &
\colhead{(s)} &
\colhead{$S/N$\tablenotemark{g}} &
\colhead{Range (\AA)} \\
\colhead{(1)} &
\colhead{(2)} &
\colhead{(3)} &
\colhead{(4)} &
\colhead{(5)} &
\colhead{(6)} &
\colhead{(7)} &
\colhead{(8)} &
\colhead{(9)} &
\colhead{(10)} &
\colhead{(11)} 
}
\startdata
$001224.03-102226.2$ & 0.2287 & 17.07 &  0.127 & --23.27 &  5 & 2010/10/12    & H2 &  1798 &   27 & 3481--5911 \\ 
                     &        &       &        &         &    & 2014/08/28    & Ab &  1560 &   41 & 2848--4476 \\ 
                     &        &       &        &         &    & 2014/08/28    & Ar &  1560 &   27 & 4197--8013 \\ 
\\
$002444.11+003221.3$ & 0.4024 & 16.85 &  0.083 & --24.86 &  4 & 2014/08/29    & Ab &  1260 &   49 & 2495--3921 \\ 
                     &        &       &        &         &    & 2014/08/29    & Ar &  1260 &   52 & 3691--7033 \\ 
\\
$015530.02-085704.0$ & 0.1648 & 16.84 &  0.080 & --22.65 &  5 & 2011/08/31\th & H2 &  2587 &   10 & 3681--6238 \\ 
                     &        &       &        &         &    & 2014/08/28    & Ab &  1260 &   28 & 3004--4721 \\ 
                     &        &       &        &         &    & 2014/08/28    & Ar &  1260 &   10 & 4426--8451 \\ 
\enddata
\tablenotetext{a}{The redshift of the object as measured in this paper
  from the peak wavelength of [\ion{O}{3}]~$\lambda$5007 line.}
\tablenotetext{b}{The apparent V magnitude, determined from the SDSS
  PSF magnitudes, as described in \S4.1 of \pone.}
\tablenotetext{c}{The Galactic visual extinction, taken from
  \citet*{schlegel98}}
\tablenotetext{d}{The absolute V magnitude, computed as described in
  \S\ref{sec:bbh_sample} of the text.}
\tablenotetext{e}{The total number of observations, including those reported in \pone.}
\tablenotetext{f}{The telescope and instrument configuration, as
  described in Table~\ref{tab:tel}.}
\tablenotetext{g}{The signal-to-noise ratio ($S/N$) in the continuum
  near the line of interest.  When the spectrum includes the H$\beta$
  line we give the $S/N$ in the continuum near this line, at 4600~\AA.
  If the spectrum includes only the \ion{Mg}{2}~$\lambda$2800 line, we
  give the $S/N$ in the continuum near this line, at 2900~\AA. Some of
  the APO spectra cover the continuum between the above lines but
  neither of the lines themselves; for these spectra we report the
  $S/N$ in the continuum at 3600~\AA.}
\tablenotetext{h}{Average of two spectra taken within a few days of
  each other. The exposure time is the sum of the individual exposure
  times and the date is the ``median'' of the dates of the two
  observations.}
\end{deluxetable*}

\begin{deluxetable*}{cllc}
\tablewidth{0in}
\tabletypesize{\scriptsize}
\tablecolumns{5}
\tablecaption{List of Telescopes and Instruments\label{tab:tel}}
\tablehead{
\colhead{Instrument} &
\colhead{} &
\colhead{} &
\colhead{Resolution\tablenotemark{a}} \\
\colhead{Code} &
\colhead{Observatory, Telescope, and Spectrograph} &
\colhead{Spectral Elements} &
\colhead{(\AA)} 
}
\startdata
MM  & MDM, Hiltner 2.4m, MODSPEC spectrograph & 600 mm$^{-1}$ grating, 1\farcs0 slit           & 3.4 \\
MO  & MDM, Hiltner 2.4m, OSMOS spectrograph & VPH grism (704 mm$^{-1}$), 1\farcs2 slit           & 3.8 \\
\\
K  & KPNO, Mayall 4m, Ritchie-Cretien spectrograph         & BL420 grating (600 mm$^{-1}$), 1\farcs5 slit &  2.4 \\
\\
Ab & APO, ARC 3.5m, Double Imaging Spectrograph            & blue arm: 400 mm$^{-1}$ grating, 1\farcs5 slit &  6.3 \\
Ar &                                                       & red arm: 300 mm$^{-1}$ grating, 1\farcs5 slit  &  7.3 \\
\\
H2 & Hobby-Eberly Telescope, Low-Resolution Spectrograph   & G2 grism (600 mm$^{-1}$), 1\farcs5 slit        &  5.6 \\
H3 &                                                       & G3 grism (600 mm$^{-1}$), 1\farcs5 slit        &  4.9 \\
\enddata
\tablenotetext{a}{The spectral resolution (FWHM) at 6400~\AA\ (the
  median wavelength of H$\beta$ for the redshift distribution of our targets).}
\end{deluxetable*}


By virtue of the sample selection, the initial spectroscopic observations come from the SDSS.  \pone\ presents additional spectra for many objects in the sample from an array of telescopes and instruments.  For this work, we use all the spectra from \pone, plus new observations that are presented here for the first time.  In Table~\ref{tab:obslog} we give a summary of the new spectroscopic observations. For each of the 212 new spectra we give the observation date, telescope and instrument used, exposure time, signal-to-noise ratio ($S/N$) in the continuum near the line of interest, and rest-frame wavelength coverage. For completeness, we also include the redshift of each object, the apparent and absolute $V$ magnitude, and the Galactic $V$-band extinction (see the footnotes to Table~\ref{tab:obslog} and section 4.1 of \pone\ for further details on the target properties) and we also list objects for which no new observations are presented. For reference, we note that in \pone\ we presented the original SDSS spectra for our 88 targets plus 108 followup spectra. We also tabulate the total number of observations per object; with the addition of the new spectra, all objects in the sample have at least two observations and most objects have at least three observations.  Time intervals between spectra in the {\it observed} frame range from weeks to over eleven years. The distribution of {\it rest}-frame time intervals is shown graphically in Figures~\ref{fig:dfdt_c} and \ref{fig:dfdt}.

\begin{figure}[t]
\epsscale{1.0} \plotone{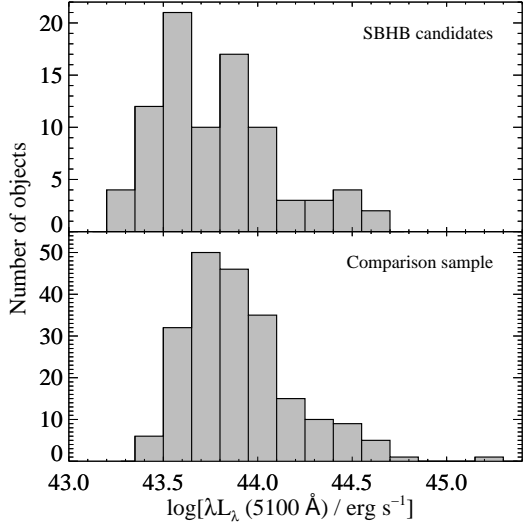} 
\caption{The distribution of continuum luminosity for the SBHB candidates and comparison sample.  Although the distributions are not identical, the samples cover a similar, relatively narrow range in luminosity which allows a reasonable comparison independent of luminosity effects.} \label{fig:Lhist}
\end{figure}

The new spectra presented here were obtained with the Hiltner 2.4m telescope at the Michigan-Dartmouth-MIT observatory (MDM), the Mayall 4m telescope at Kitt Peak National Observatory (KPNO), the Astrophysical Research Consortium (ARC) 3.5m telescope at Apache Point Observatory (APO), and the 9.2m Hobby-Eberly Telescope (HET). The instrument configurations used to obtain the spectra are listed in Table~\ref{tab:tel} along with the spectral resolution attained by each instrument configuration. The spectra were reduced and calibrated as described in section 5.1 of \pone. Since accurate wavelength calibration is important for our purposes, we summarize the wavelength calibration process here. We used 20--60 arc emission lines to derive the wavelength solution for each observing run. This solution consists of a polynomial of order five or less, connecting the detector pixel number to the wavelength. We required that the standard deviation of fit residuals was 0.1 pixel or less. Thus, the {\it relative} velocity scale of each spectrum at 6400~\AA\ (the median wavelength of H$\beta$ for our quasars) is good to 5~\kms\ for MDM spectra, 7~\kms\ for KPNO spectra, 11~\kms\ for APO spectra, and 9~\kms\ for HET spectra. We used the \flion{O}{1}{5577} line in the night-sky spectrum, recorded simultaneously with and on the same detector as the quasar spectrum, to set the {\it absolute} wavelength scale of our new spectra. After applying heliocentric velocity corrections and converting the wavelength scale to vacuum values, we aligned the quasar's \fllion{O}{3}{4959,~5007} doublet in each new spectrum to that of the original SDSS spectrum using the cross-correlation method described in \pone. The final uncertainty in this alignment translates to an uncertainty in relative velocities between the broad H$\beta$ lines observed at different epochs. As noted in \pone, this uncertainty is smaller than 25~\kms\ for all objects and smaller than 7~\kms\ in 80\% of objects in our sample.

\begin{figure}[t]
\epsscale{1.3} \plotone{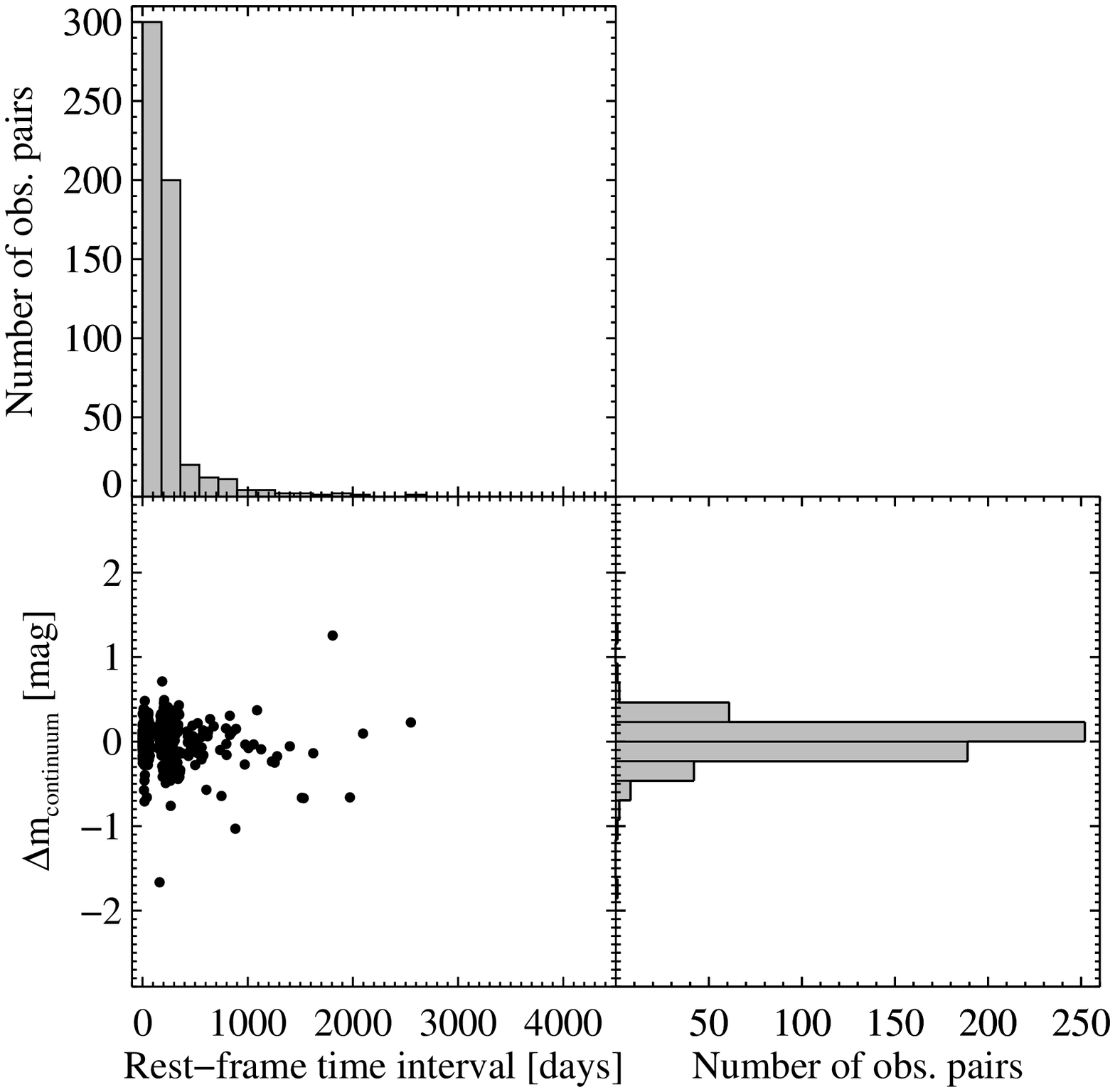} 
\caption{For every unique pair of observations for the comparison sample, the distribution of rest-frame time intervals (top), $\Delta m=-2.5\textrm{log}(f_1/f_2)$ measured from continuum flux density ratios at 5100~\AA\ (bottom right), and $\Delta m$ at 5100~\AA\ as a function of rest-frame time interval between observations (bottom left).  The $\Delta m$ distribution has a median of 0.03 and a standard deviation of 0.22.} \label{fig:dfdt_c_sdss}
\vfill
\epsscale{1.3} \plotone{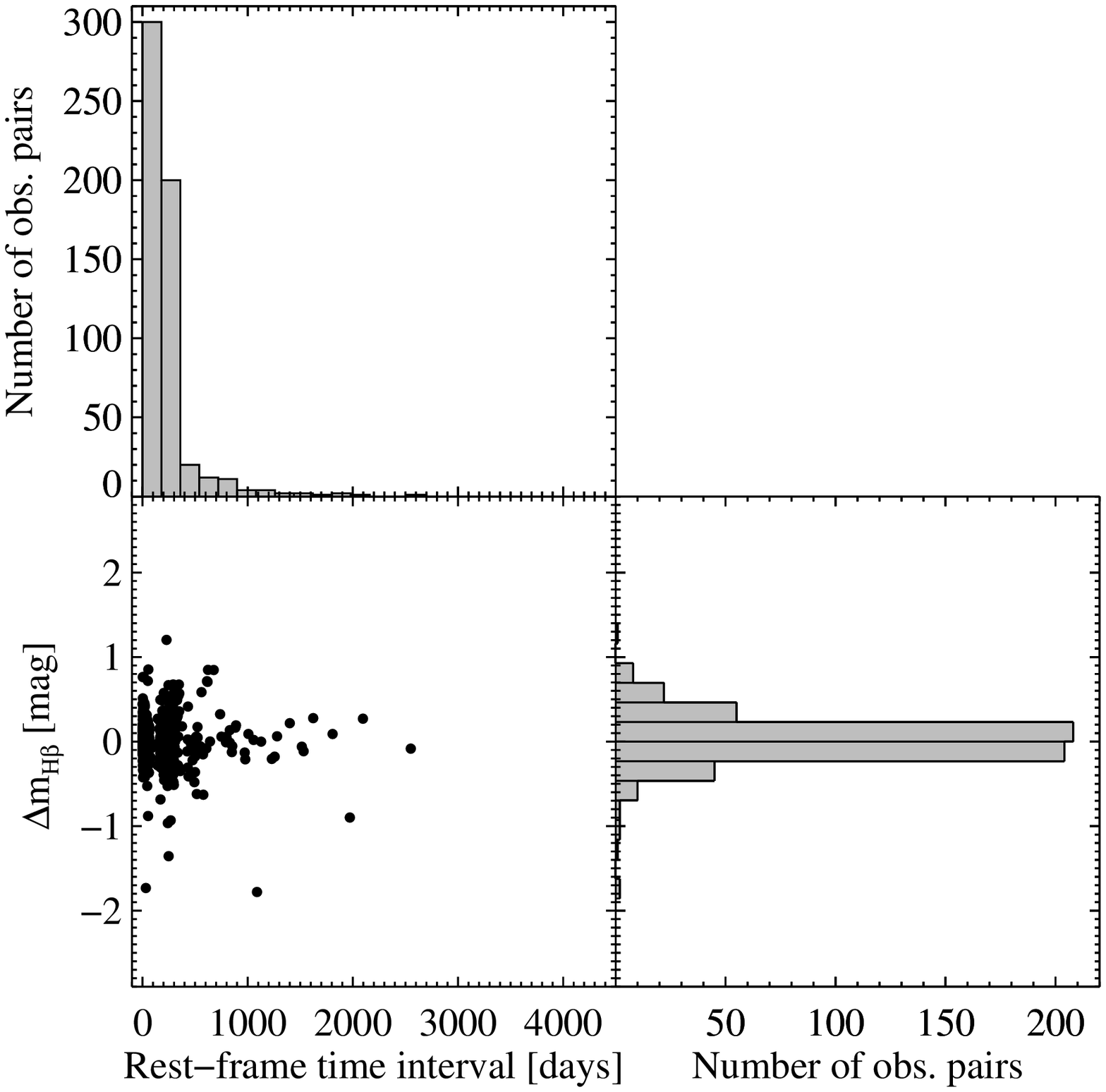} 
\caption{For every unique pair of observations for the comparison sample, the distribution of rest-frame time intervals (top), $\Delta m=-2.5\textrm{log}(f_1/f_2)$ measured from integrated \Hb\ flux ratios (bottom right), and the integrated \Hb\ $\Delta m$ as a function of rest-frame time interval between observations (bottom left).  The $\Delta m$ distribution has a median of 0.02 and a standard deviation of 0.28.} \label{fig:dfdt_sdss}
\end{figure}

\subsection{A Comparison Sample of Typical SDSS Quasars}
\label{sec:sdss_sample}
In order to characterize the variability properties of typical quasars, we generated a comparison sample of SDSS quasars with multiple spectroscopic observations in the seventh data release (DR7) quasar catalog \citep{schneider07}.  Typically, each object is observed two to three times, although several objects have as many as ten observations.  Starting with a list of objects with multiple observations, we selected those in the redshift range $0.08<z<0.7$, which matches that of our SBHB candidates. To ensure our ability to make reliable measurements from the spectra, we also required that $S/N > 20$ in the \Hb\ region \citep[using the $S/N$ values from][]{shen11}.  These criteria resulted in a comparison sample of 212 quasars that we analyze in an identical way to the sample of SBHB candidates throughout this work to facilitate a fair comparison free of any methodological biases.

All observations of quasars in the comparison sample were carried out with the SDSS spectrograph through fibers of diameter 3\arcsec. The resulting spectral resolution was 2.7~\AA, which is slightly better than that achieved in followup observations of the sample of SBHB candidates (see Table~\ref{tab:tel}). Differences in spectral resolution between SDSS spectra and followup spectra is inconsequential for our purposes since all the lines of interest, including the narrow H$\beta$ and \fion{O}{3} lines, are always broader than the instrumental resolution.

Given that quasar variability properties are known to depend on luminosity and redshift, it is important that the redshifts and luminosities (primarily the luminosities) of the comparison sample span the same range as those of the SBHB candidates. The redshift distribution of the quasars in the comparison sample is shown in the bottom panel of Figure~\ref{fig:zhist}.  The redshifts span the range $0.08<z<0.68$, with a median value of 0.38, very similar to the binary candidates described in Section~\ref{sec:bbh_sample}.  The monochromatic luminosity distributions of the binary candidates and the comparison sample are shown in Figure~\ref{fig:Lhist} (see Section~\ref{sec:sfit} below for a description of the luminosity measurements).  The binary candidates span the range $43.27 < \textrm{log}\,\left[\lambda\,L_{\lambda}(5100\,\textrm{\AA})/\textrm{erg~s}^{-1}\right]  <44.68$ with a median value of 43.75 and a standard deviation of 0.31.  The comparison sample spans the range $43.44 < \textrm{log}\,\left[\lambda\,L_{\lambda}(5100\,\textrm{\AA})/\textrm{erg~s}^{-1}\right]  <45.17$ with a median value of 43.88 and a standard deviation of 0.33. The shapes of the redshift and luminosity distributions of the SBHB candidates and comparison quasars are not identical, of course, as indicated by the Kolmolgorov-Smirnov (KS) test, which gives probabilities that the redshift and luminosity distributions were drawn from the same parent population of 0.002 and 0.006, respectively.

The distribution of rest-frame time intervals is shown in the top panels of both Figure~\ref{fig:dfdt_c_sdss} and Figure~\ref{fig:dfdt_sdss}.  The median, observed-frame time interval between observations of the same object in this sample is approximately 225~days, with concentrations of time interval between observations at less than 100~days and approximately 300~days (these are not obvious in relevant histograms because a very fine binning would be required to resolve them, but the concentrations are evident in the associated scatter plots).  The minimum time interval in the observed frame is less than a day, and the maximum is nearly eight years.

\subsection{Spectral Decomposition and Emission-Line Measurements}
\label{sec:sfit}
In order to isolate the broad \Hb\ line and quasar continuum, we perform a spectral decomposition that deblends the quasar continuum, optical \FeII, \Hb, and \OIII\ emission components.  All line and continuum fitting is performed using the IRAF task \textsc{specfit} \citep{kriss94} in three steps.  This process is qualitatively similar to the approach of \pone\ and was adopted because, in objects where the line profiles are particularly complex, it is more effective at characterizing the red wing of \Hb\ under the \OIII\ emission than methods that fit all the components simultaneously \citep[e.g.,][]{shang05}.  In fact, attributing various components of the fit to different physical regions is not advisable, but it is unavoidable in this case, and we feel that this method minimizes the danger of misrepresenting the data.
  
In the first step, we fit and subtract the optical continuum, which includes contributions that can be characterized by the combination of a featureless power law of the form $f_{\lambda} \propto \lambda^{-\alpha}$ and the optical \FeII\ template of \citet{veron-cetty04}.  We adopt two suites of two initial guesses, where the initial guesses differ in the properties of the starting power law.  The first suite includes only the power-law continuum and \FeII\ template.  The second suite adds a Gaussian to account for \HeIIw\ emission and was only selected when the \HeII\ emission was visually evident in the spectrum.  The fit is optimized in three windows at 4150--4200, 4435--4700, and 5100--5700~\AA\ and the best fit is selected by applying the Bayesian Information Criterion \citep[hereafter BIC,][]{schwartz78}.  Although it is appropriate to minimize the $\chi^2$ statistic when performing the fit, the reduced $\chi^2$ is ineffective at choosing between models because a fit with more free parameters will always perform better at this test.  The BIC provides an evaluation of the goodness of fit, similar to that provided by the reduced $\chi^2$, but penalizes the model based on the number of free parameters, and thus evaluates whether the addition of new free parameters truly improves the fit.  In a minority of cases, the optical continuum was not well fitted by the above procedure; For example in some cases where data were missing in one of the fit windows.  In these cases, manual adjustments were made to the initial guess or wavelength range of the fit.  Finally, continuum-subtracted spectra were generated by subtracting the best-fitting optical continuum model from the data.

\begin{figure}[t]
\epsscale{1.1} \plotone{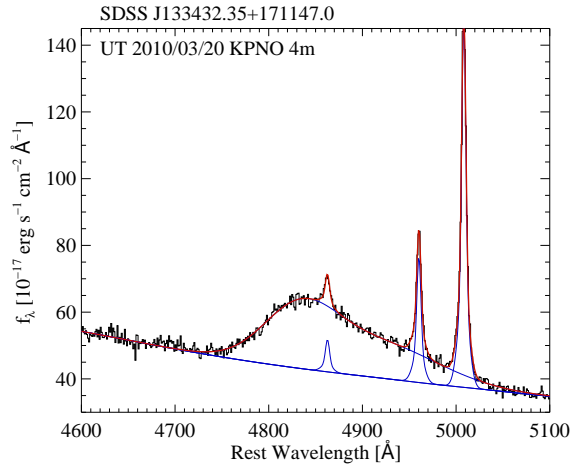} 
\caption{An example of the spectral decomposition for SDSS J133432.35$+$171147.0.  The red line shows the total model, including the power-law continuum, \FeII\ template (which is weak in this object), the narrow \OIII\ lines, and the broad and narrow \Hb\ lines.  The individual line profiles are shown in blue.} \label{fig:specfit}
\end{figure}

In the second step, we characterize the \OIIIdblt\ emission lines and subtract them.  We parameterize the profile of narrow emission lines in the spectrum based on the profile of the \OIIIw\ line. In order to isolate that line, we fit a low-order polynomial to the local continuum (i.e. the red wing of the broad \Hb\ line) in manually selected wavelength windows.  The polynomial is temporarily subtracted and the remaining  \OIIIw\ line is fitted with two Gaussians to adequately account for any possible profile asymmetry (no physical meaning is attributed to the individual Gaussians).  The \flion{O}{3}{4959} line profile is then taken to be identical to that of \OIIIw, but appropriately shifted in wavelength and scaled down in flux by a factor of $1/3$.  With the \OIII\ profiles thus characterized, the low-order polynomial is added back to the data and the \OIII\ doublet is subtracted to isolate the \Hb\ profile.

In the third and final step, we decompose the \Hb\ profile.  In typical quasars, the \Hb\ profile can be adequately represented by a combination of four Gaussian components.  Two of the Gaussians account for the narrow \Hb\ line and are constrained to create a profile identical in shape and tied in wavelength to the \OIIIw\ line, although the flux of the profile is allowed to vary and even to be zero if this provides the best fit.  The other two Gaussians are allowed to vary freely to characterize the broad \Hb\ line.  In cases where these profiles are very complex, an additional Gaussian is necessary to adequately characterize them.  These cases are identified by visual inspection, and are outliers in the distribution of reduced $\chi^2$ when we fit them with only two Gaussians (typically, $\chi^2/\nu>16$).  Including the third Gaussian results in a decrease in the BIC of $>10$, indicating that the additional free parameters associated with the third Gaussian significantly improve the fit.  An example of the fitting results is shown in Figure~\ref{fig:specfit}.

We measure spectral properties from the parametric models obtained for the \Hb\ and \OIIIw\ lines in the SDSS spectra and give the results in Tables~\ref{tab:hbmeasurements} and \ref{tab:o3measurements}. The total line fluxes are obtained by integrating the emission-line model where the model is above 1\% of the peak flux density of the line.  We verify visually that this scheme includes the entire line flux (see Appendix~\ref{app:bwin} for a detailed discussion). The continuum luminosity is measured at 5100~\AA\ from the power-law component of the continuum fit and is included in Table~\ref{tab:o3measurements}. The continuum luminosity distributions for the SBHB candidates and comparison sample are shown in Figure~\ref{fig:Lhist}.  The luminosity range in both samples is relatively narrow, spanning only about an order of magnitude.

Using the spectral decompositions, we also determine a suite of parameters to characterize the  \Hb\ and \OIIIw\ line profiles. Although not all of these quantities are needed for the variability analysis carried out in this paper, we include them in Tables~\ref{tab:hbmeasurements} and \ref{tab:o3measurements} for future reference.  Following \pone, the first four central moments of the line profiles, $\mu_{n}$, are calculated from $\mu_{n} = K \sum\limits_{i} (\lambda_{i} - \langle\lambda\rangle)^{n} f_{i}$, where $\lambda_{i}$ and $f_{i}$ are the discrete wavelength and flux densities, the normalization constant $K$ is given by $1/K = \sum\limits_{i}f_{i}$, and $\langle\lambda\rangle = K\sum\limits_{i}\lambda_{i}f_{i}$ is the first moment (i.e. the centroid) of the line profile.  From these measurements we can derive the skewness coefficient, $s = \mu_{3}/\mu_{2}^{3/2}$, which describes the degree of asymmetry of the profile.  Symmetric profiles have $s=0$ and positive values of the skewness coefficient indicate that the line leans to bluer wavelengths (i.e., the red wing is more extended than the blue).  The kurtosis coefficient, $k = \mu_{4}/\mu_{2}^2$, describes the shape of the line profile, with boxy profiles having smaller values of $k$.  We also calculate the Pearson skewness coefficient, $p = (\langle\lambda\rangle-\lambda_{m})/\mu_{2}^{1/2}$.  In this case, $\lambda_{m}$ is the median wavelength corresponding to the point where the area of the line profile is split in half.  Note that this definition matches that given in \pone, but the values given in Table~4 of \pone\ had the sign reversed, inadvertently. We correct this mistake in this paper. Positive values of the Pearson skewness coefficient also describe blue-leaning line profiles.

\begin{figure}[t]
\epsscale{1.0} \plotone{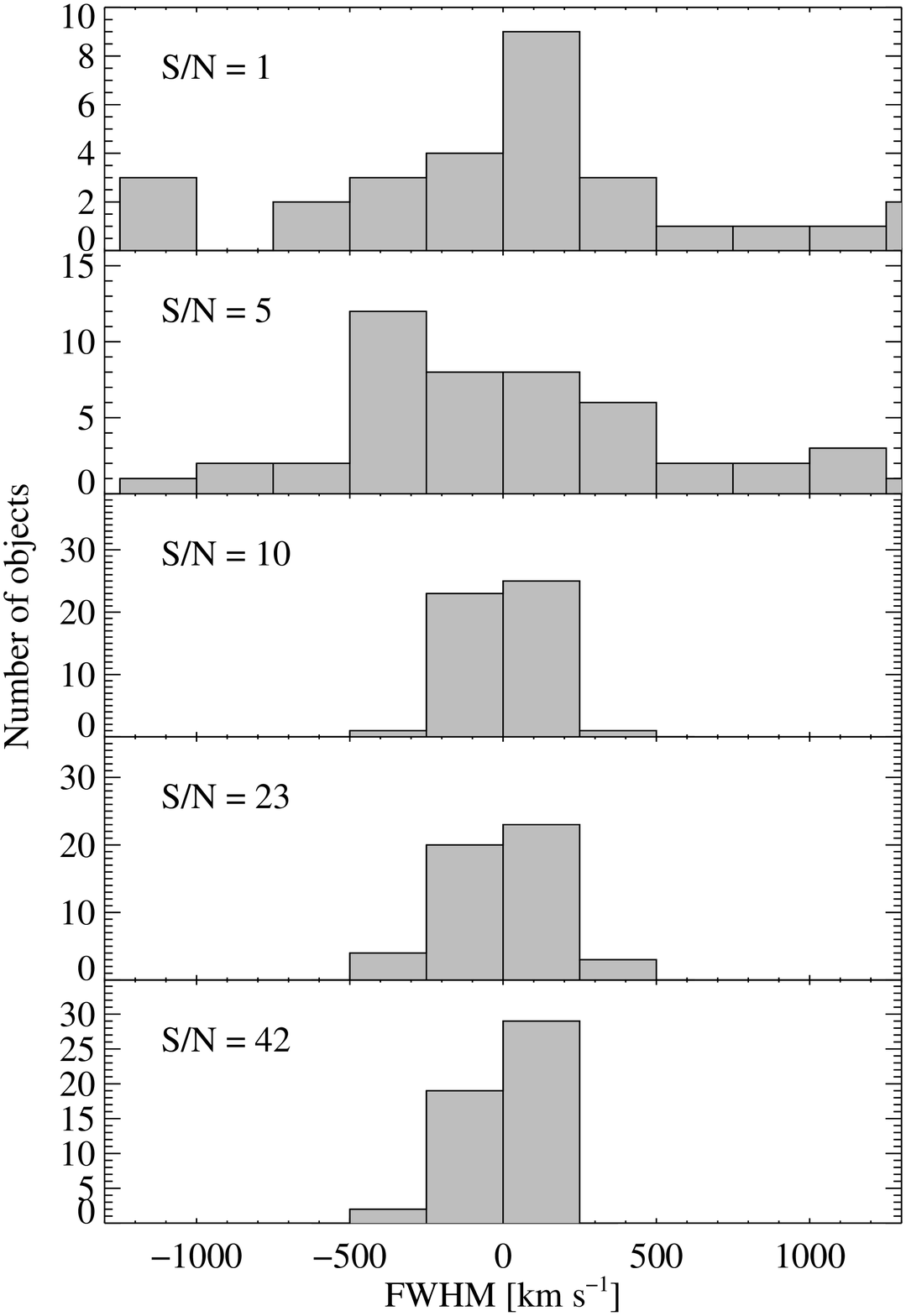} 
\caption{Distributions of the FWHM of \Hb\ measured from the 50 synthetic spectra generated for 5 different objects.  Each panel represents a different object and is labeled with the value of $S/N$ measured from the observed spectrum.  The standard deviations of these distributions yield the FWHM uncertainties as a function of $S/N$.\label{fig:synthdist}}
%
\epsscale{1.0} \plotone{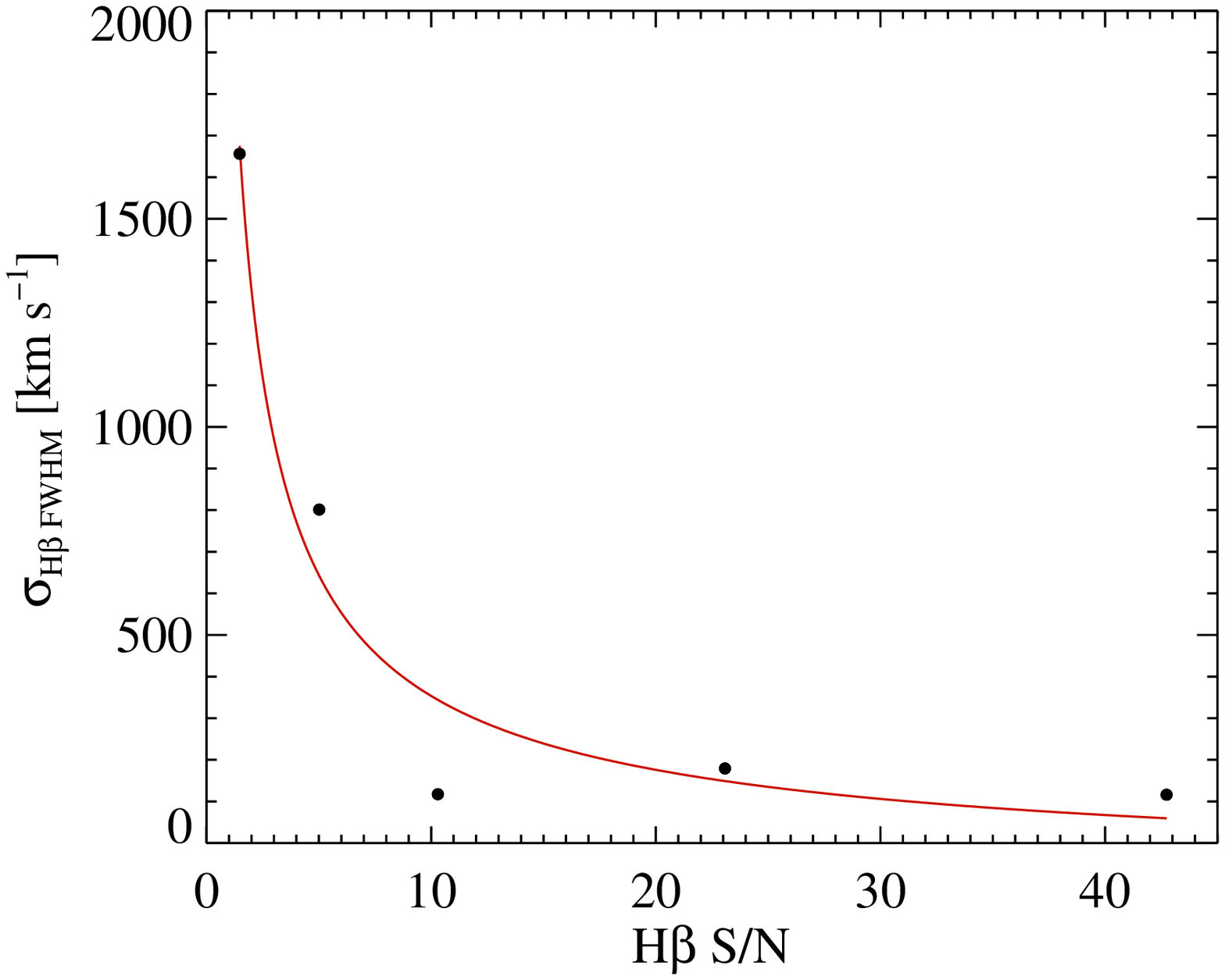} 
\caption{Uncertainties in the FWHM taken from the widths of the distributions in the previous figure versus $S/N$.  These are fitted with a function of the form $\sigma=A\,(S/N)^{-\alpha}+B$, where $A=2341.33$, $B=-103.40$, and $\alpha=0.71$ in this case.\label{fig:synthfit}} 
\end{figure}

An accurate accounting of the uncertainties associated with the spectral properties must include both the effect of noise in the spectrum as well as uncertainties incurred during the interactive fitting procedure.  To estimate these uncertainties we carry out Monte Carlo simulations of the spectral decomposition.  First, we select five spectra that span the range of $S/N$ present in the sample of SBHB candidates, making an effort to include instances of qualitatively different spectral properties.  We calculate the $S/N$ within $\pm25$~\AA\ of the peak of the broad \Hb\ profile after continuum subtraction.  For each of these five objects, we generate fifty realizations of the best-fitting model spectrum by adding Gaussian noise with a constant amplitude equal to that measured at the peak of the broad \Hb\ line.  Each synthetic spectrum is then run through the spectral fitting procedure (including the interactive continuum placement and line fitting) and all of the spectral properties are measured by fitting the synthetic spectrum in the same manner as the observed spectrum (see description above). We adopt the resulting standard deviation in the distribution of each measured property as the uncertainty for that measurement at that $S/N$.  As an example of the outcome of this experiment, we show in Figure~\ref{fig:synthdist} the distribution of \Hb\ FWHM values for different values of the $S/N$. Next, we fit a simple function of the form $\sigma = A\,(S/N)^{-\alpha}+B$ to the relation between the $S/N$ and corresponding uncertainty, as seen in the example of Figure~\ref{fig:synthfit}.  We find that the uncertainty usually scales as either $(S/N)^{-1}$ or $(S/N)^{-1/2}$.  In a handful of cases, we fit the fractional uncertainty (relative to the mean of the distribution of realizations) rather than the absolute uncertainty as a function of $S/N$ because we find that this relationship is much more clear.  Using these relationships we estimate uncertainties on the measured quantities for the objects in the sample.  For the FWHM and peak wavelength shift of \OIII, we find a negligible spread in the values measured from the synthetic spectra, likely because the line profiles have strong peaks and few data points, therefore we set the following upper limits to these uncertainties: 70~km~s$^{-1}$ and 30~km~s$^{-1}$, respectively. In Appendix~\ref{app:measurements} we investigate the sources of uncertainty in detail, including the uncertainty arising from the choice of method for decomposing the spectra. The results presented in the tables of this paper do not include the uncertainty from the choice of decomposition method.

\begin{deluxetable*}{ccccccccccc}
\setlength{\tabcolsep}{-3pt}
\tabletypesize{\scriptsize}
\tablewidth{0in}
\tablecaption{Spectral measurements for broad H$\beta$ [Abridged]
\label{tab:hbmeasurements}
}
\tablehead{
\colhead{} & 
\colhead{} & 
\colhead{} & 
\colhead{} & 
\colhead{Peak} & 
\colhead{Centroid} & 
\colhead{} & 
\colhead{} & 
\colhead{} & 
\colhead{} & 
\colhead{} \\ 
\colhead{Object} & 
\colhead{Obsev.} & 
\colhead{} & 
\colhead{} & 
\colhead{Velocity} & 
\colhead{Velocity} & 
\colhead{Velocity} & 
\colhead{Pearson} & 
\colhead{} & 
\colhead{} & 
\colhead{} \\ 
\colhead{Name} & 
\colhead{Date} & 
\colhead{FWHM} & 
\colhead{FWQM} & 
\colhead{Shift} & 
\colhead{Shift} & 
\colhead{Dispersion} & 
\colhead{Skewness} & 
\colhead{Kurtosis} & 
\colhead{Integr.} & 
\colhead{EW} \\ 
\colhead{SDSSJ} & 
\colhead{(UT)} & 
\colhead{(km/s)} & 
\colhead{(km/s)} & 
\colhead{(km/s)} & 
\colhead{(km/s)} & 
\colhead{(km/s)} & 
\colhead{Coeff.} & 
\colhead{Coeff.} & 
\colhead{Flux\,\tablenotemark{a}} & 
\colhead{(\AA)} \\ 
\colhead{(1)} & 
\colhead{(2)} & 
\colhead{(3)} & 
\colhead{(4)} & 
\colhead{(5)} & 
\colhead{(6)} & 
\colhead{(7)} & 
\colhead{(8)} & 
\colhead{(9)} & 
\colhead{(10)} & 
\colhead{(11)} 
}
\startdata
001224
&
2001/08/20
&
3200
$\pm$
300
&
9300
$\pm$
400
&
$-$1880
$\pm$
90
&
900
$\pm$
200
&
4200
$\pm$
200
&
0.23
$\pm$
0.02
&
    2.86
$\pm$
    0.04
&
3500
$\pm$
100
&
92
$\pm$
4
\\
 & 
2009/12/16
&
2600
$\pm$
300
&
8700
$\pm$
400
&
$-$2080
$\pm$
80
&
6
$\pm$
200
&
3900
$\pm$
200
&
0.26
$\pm$
0.02
&
    3.13
$\pm$
    0.04
&
3200
$\pm$
100
&
104
$\pm$
4
\\
 & 
2010/10/12
&
4400
$\pm$
200
&
8100
$\pm$
300
&
$-$2120
$\pm$
60
&
200
$\pm$
200
&
3900
$\pm$
200
&
0.19
$\pm$
0.01
&
    3.71
$\pm$
    0.04
&
9000
$\pm$
200
&
98
$\pm$
4
\\
 & 
2014/08/28
&
4600
$\pm$
300
&
9000
$\pm$
400
&
$-$2030
$\pm$
90
&
400
$\pm$
200
&
4200
$\pm$
200
&
0.20
$\pm$
0.02
&
    3.52
$\pm$
    0.04
&
8300
$\pm$
300
&
134
$\pm$
6
\\
\\
002444
&
2000/12/22
&
8490
$\pm$
100
&
       13200
$\pm$
200
&
$-$170
$\pm$
40
&
100
$\pm$
90
&
4230
$\pm$
70
&
0.030
$\pm$
0.008
&
    3.01
$\pm$
    0.03
&
4010
$\pm$
80
&
93
$\pm$
3
\\
 & 
2009/12/18
&
9400
$\pm$
200
&
       14800
$\pm$
400
&
$-$410
$\pm$
80
&
$-$400
$\pm$
200
&
4800
$\pm$
200
&
0.01
$\pm$
0.01
&
    3.14
$\pm$
    0.04
&
1840
$\pm$
50
&
106
$\pm$
4
\\
 & 
2014/08/29
&
       10100
$\pm$
200
&
       17400
$\pm$
300
&
$-$330
$\pm$
60
&
1900
$\pm$
100
&
6600
$\pm$
100
&
0.155
$\pm$
0.009
&
    3.64
$\pm$
    0.03
&
5400
$\pm$
100
&
136
$\pm$
5
\\
\\
015530
&
2001/09/16
&
7500
$\pm$
200
&
       10500
$\pm$
300
&
2120
$\pm$
70
&
1100
$\pm$
200
&
3500
$\pm$
200
&
0.04
$\pm$
0.01
&
    3.43
$\pm$
    0.04
&
6300
$\pm$
200
&
108
$\pm$
4
\\
 & 
2009/12/17
&
7200
$\pm$
200
&
       10100
$\pm$
300
&
2200
$\pm$
70
&
1000
$\pm$
200
&
3400
$\pm$
200
&
0.07
$\pm$
0.01
&
    3.32
$\pm$
    0.04
&
2460
$\pm$
70
&
98
$\pm$
4
\\
 & 
2011/08/31
&
7100
$\pm$
200
&
       10000
$\pm$
300
&
2320
$\pm$
60
&
600
$\pm$
200
&
3000
$\pm$
200
&
$-$0.01
$\pm$
0.01
&
    2.86
$\pm$
    0.04
&
4300
$\pm$
100
&
77
$\pm$
3
\\
 & 
2014/08/28
&
6400
$\pm$
300
&
       11500
$\pm$
500
&
40
$\pm$
100
&
800
$\pm$
200
&
4400
$\pm$
300
&
0.06
$\pm$
0.05
&
    3.55
$\pm$
    0.05
&
7100
$\pm$
300
&
103
$\pm$
5
\enddata
\tablenotetext{a}{Integrated fluxes are in units of $10^{-17}$ erg s$^{-1}$ cm$^{-2}$.}
\end{deluxetable*}

\begin{deluxetable*}{cccccccccccc}
\setlength{\tabcolsep}{1pt}
\tabletypesize{\scriptsize}
\tablecolumns{9}
\tablewidth{0in}
\tablecaption{Spectral measurements for [O {\sc iii}]~$\lambda5007$ [Abridged]
\label{tab:o3measurements}
}
\tablehead{
\colhead{} & 
\colhead{} & 
\colhead{} & 
\colhead{} & 
\colhead{} & 
\colhead{} & 
\colhead{} & 
\colhead{} & 
\colhead{} \\ 
\colhead{Object} & 
\colhead{} & 
\colhead{} & 
\colhead{} & 
\colhead{Velocity} & 
\colhead{Pearson} & 
\colhead{} & 
\colhead{} & 
\colhead{} \\ 
\colhead{Name} & 
\colhead{Observation} & 
\colhead{} & 
\colhead{FWHM\,\tablenotemark{b}} & 
\colhead{Dispersion} & 
\colhead{Skewness} & 
\colhead{Kurtosis} & 
\colhead{Integrated} & 
\colhead{EW} \\ 
\colhead{SDSS J} & 
\colhead{Date (UT)} & 
\colhead{$f_{\lambda}$\,(5100\,\AA)\,\tablenotemark{a}} & 
\colhead{(km s$^{-1}$)} & 
\colhead{(km s$^{-1}$)} & 
\colhead{Coefficient} & 
\colhead{Coefficient} & 
\colhead{Flux\,\tablenotemark{c}} & 
\colhead{(\AA)} \\ 
\colhead{(1)} & 
\colhead{(2)} & 
\colhead{(3)} & 
\colhead{(4)} & 
\colhead{(5)} & 
\colhead{(6)} & 
\colhead{(7)} & 
\colhead{(8)} & 
\colhead{(9)} 
}
\startdata
001224
&
2001/08/20
&
36.0
$\pm$
0.2
&
620
&
310
$\pm$
20
&
$-$0.20
$\pm$
0.04
&
    3.17
$\pm$
    0.05
&
360
$\pm$
20
&
10.0
$\pm$
0.6
\\
 & 
2009/12/16
&
29.0
$\pm$
0.2
&
750
&
510
$\pm$
20
&
$-$0.14
$\pm$
0.04
&
    3.44
$\pm$
    0.05
&
460
$\pm$
20
&
16.0
$\pm$
0.8
\\
 & 
2010/10/12
&
84.0
$\pm$
0.2
&
680
&
570
$\pm$
10
&
$-$0.07
$\pm$
0.03
&
    3.80
$\pm$
    0.04
&
1590
$\pm$
50
&
18.0
$\pm$
0.7
\\
 & 
2014/08/28
&
59.0
$\pm$
0.2
&
780
&
780
$\pm$
20
&
0.08
$\pm$
0.04
&
    3.83
$\pm$
    0.05
&
1360
$\pm$
70
&
23
$\pm$
1
\\
\\
002444
&
2000/12/22
&
39.0
$\pm$
0.2
&
410
&
298
$\pm$
8
&
$-$0.06
$\pm$
0.02
&
    4.12
$\pm$
    0.02
&
544
$\pm$
9
&
13.0
$\pm$
0.3
\\
 & 
2009/12/18
&
16.0
$\pm$
0.2
&
520
&
390
$\pm$
20
&
$-$0.05
$\pm$
0.03
&
    4.82
$\pm$
    0.04
&
260
$\pm$
10
&
16.0
$\pm$
0.8
\\
 & 
2014/08/29
&
37.0
$\pm$
0.2
&
490
&
340
$\pm$
10
&
0.18
$\pm$
0.03
&
    4.53
$\pm$
    0.03
&
510
$\pm$
10
&
13.0
$\pm$
0.4
\\
\\
015530
&
2001/09/16
&
54.0
$\pm$
0.2
&
280
&
160
$\pm$
10
&
$-$0.06
$\pm$
0.03
&
    3.69
$\pm$
    0.04
&
1460
$\pm$
60
&
26
$\pm$
1
\\
 & 
2009/12/17
&
23.0
$\pm$
0.2
&
470
&
250
$\pm$
10
&
$-$0.28
$\pm$
0.03
&
    3.75
$\pm$
    0.04
&
870
$\pm$
30
&
36
$\pm$
1
\\
 & 
2011/08/31
&
52.0
$\pm$
0.2
&
310
&
230
$\pm$
10
&
$-$0.16
$\pm$
0.03
&
    4.40
$\pm$
    0.04
&
1310
$\pm$
40
&
25
$\pm$
1
\\
 & 
2014/08/28
&
65.0
$\pm$
0.2
&
470
&
200
$\pm$
30
&
0.08
$\pm$
0.04
&
    3.19
$\pm$
    0.06
&
1800
$\pm$
100
&
28
$\pm$
2
\enddata
\tablenotetext{a}{Flux densities are in units of $10^{-17}$ erg s$^{-1}$ cm$^{-2}$ \AA$^{-1}$.}
\tablenotetext{b}{Uncertainties on the \OIII\ FWHM are less than 70~km~s$^{-1}$.}
\tablenotetext{c}{Integrated fluxes are in units of $10^{-17}$ erg s$^{-1}$ cm$^{-2}$.}
\end{deluxetable*}


The primary objective of the multi-epoch spectroscopy of the SBHB candidates is to search for radial velocity variations, which will be the topic of an upcoming paper.  As a result, while particular care was taken with the wavelength calibration, the {\it absolute} flux calibration can be uncertain by up to a factor of two, largely because of variable seeing.  To bypass this uncertainty, we assume that the integrated \OIII\ luminosity does not change over the time scales spanned by our observations.  To support this assumption, we verify that the narrow-line regions (NLRs) of our targets have a smaller angular size than the spectrograph aperture (circular fibers with a radius of 1\farcs5 in the case of SDSS spectra and a slit with a width of 1\farcs5 in the case of follow-up spectra).  To do this, we use luminosity-based prescription of \citet{bennert02}, which gives a measure of the outer extent of the NLR ($R_{NLR}$, defined as the radius of the ring aperture where the \OIII\ flux in their images drops to 3$\sigma$ above the background and typically enclosing 98\% of the detectable emission) as a function of the \OIII\ luminosity.  The scatter in this relationship is 0.14~dex in log($R_{NLR}$), so in order to obtain an upper limit on the size of the NLR in our objects, we add this to the values calculated from the prescription.  For maximum sizes of the NLRs in our sample, we use the \OIII\ luminosity from the SDSS spectrum for each object (because these have the best flux calibration) and obtain $800\,\textrm{pc} < R_{NLR,\, max} < 14600\,\textrm{pc}$ with a median value of 2700~pc and corresponding angular sizes of $0\farcs12 < \alpha < 1\farcs46$ with a median value of 0\farcs42.  Based on this calculation, none of the objects in the sample have maximum NLR sizes larger than 1\farcs5.  The angular sizes that we estimate are also consistent with expectations from more recent work \citep[e.g.,][]{hainline13,hainline14}.  Given that the images of \citet{bennert02} and \citet{young14} show that the NLR emission is actually concentrated far inside of the outermost extent that is measured by $R_{NLR}$, we determine that there should be no variability as a result of a small size or orientation of the spectroscopic slit.  Intrinsic narrow-line variability in response to changes in the ionizing continuum can take place on time scales of years \citep[e.g., NGC~5548 which demonstrates about 20\% variation over 20 years, suggesting that the NLR extends down to a few parsecs,][]{peterson13}.  Because the NLR is so large in the binary candidates and comparison sample, which are more luminous than NGC~5548, we do not expect intrinsic variability on the time scales of our observations.  Thus, whenever we compare two spectra from this point forward, we scale all flux densities and integrated line fluxes so that the integrated \OIII\ flux of the later observation matches that of the earlier one.

\section{Analysis}
\label{sec:analysis}

\subsection{The Distribution of Magnitude Differences}
In this work, we consider the rest-frame, ensemble flux variability in the continuum and \Hb\ emission line.  Because each object in the SBHB candidate sample has only been observed a handful of times, the light curves are poorly sampled and cannot provide a useful characterization of the variability of individual objects.  Therefore, we consider the variability of the sample of SBHB candidates as a whole by means of the structure function.  In this approach, we consider every possible pair of spectra for each object to obtain a measure of the dispersion in the magnitudes, where $\Delta m=-2.5\, \log(f_1/f_2)$, at different time scales for the entire sample.

For the SBHB candidates, the right panels of Figures~\ref{fig:dfdt_c} and \ref{fig:dfdt} show the distributions of continuum and integrated \Hb\ magnitude differences, respectively, of unique pairs of observations.  Although the distribution is centered near zero there are noticeable wings that are more pronounced for \Hb\ than the continuum. The distribution of continuum magnitude differences has a median of $-0.05$ and standard deviation of 0.43 while the distribution of \Hb\ magnitude differences has a median of $-0.03$ and a standard deviation of 0.59.

The distribution of magnitudes does not take into account the time interval between observations, which is known to play a role in determining the amplitude of flux variability in typical quasars \citep{vandenberk04,macleod12}.  Figures~\ref{fig:dfdt_c} and \ref{fig:dfdt} also show the continuum and integrated \Hb\ magnitude differences, respectively, for the SBHB candidates as a function of the rest-frame time delay between the observations. It is readily apparent that a substantial fraction of observations are separated by 1000--3000~days. This is a direct result of the observing strategy we adopted, where the goal was to evaluate the binary hypothesis based on the \Hb\ radial velocity variability. It is also noteworthy that the distributions of continuum and integrated \Hb\ magnitude differences are fairly similar. 

\subsection{Ensemble Variability}
\label{sec:sf}
In order to quantify the ensemble variability of the SBHB candidates as a function of the time interval between observations, we compute the structure function, which provides a measure of the average relative change in the flux of the objects in the sample. There are multiple definitions in the literature, we adopt two here.  The first version and the corresponding uncertainty, from \citet{vandenberk04}, are defined as follows:
\begin{eqnarray}
\label{eqn:v04}
SF (\Delta t)&=& \sqrt{\frac{\pi}{2}\langle |\Delta m|\rangle^2-\langle \sigma_{\Delta m}^2\rangle} \\
\sigma_{SF} &=& \frac{1}{2\, SF} \sqrt{\pi^2 \langle | \Delta m| \rangle^2 \sigma_{rms,\Delta m}^2 + \sigma_{rms,\sigma_{\Delta m}}^2 }.
\end{eqnarray}

We evaluate the magnitude difference $\Delta m=-2.5\, \log(f_1/f_2)$ between two epochs (rest-frame time interval $\Delta t$), after scaling the later spectrum to the same \OIII\ flux as as the former spectrum. The quantity $\sigma_{\Delta m}$ represents the formal measurement uncertainty in $\Delta m$ and $\sigma_{rms,\Delta m}$ and $\sigma_{rms,\sigma_{\Delta m}}$ are the standard deviations of the $\Delta m$ and $\sigma_{\Delta m}$ distributions.  The structure function is then calculated in time interval bins.  

The second version of the structure function we consider is defined by \citet{macleod12} as
\begin{equation}
\label{eqn:viqr}
SF_{IQR}(\Delta t) = \sqrt{[0.74\,IQR(\Delta m)]^2 - \widetilde{\sigma_{\Delta m}^2}},
\end{equation}
where $IQR$ is the interquartile range (full width at quarter maximum) of the cumulative $\Delta m$ distribution and $\widetilde{\sigma_{\Delta m}^2}$ is the median of the squared $\Delta m$ measurement uncertainties.  The uncertainty in this structure function is simulated by bootstrap resampling of the distribution of $\Delta m$ 1000 times, calculating the corresponding value of the interquartile range for each iteration, and finally evaluating the width of the resulting distribution of $SF_{IQR}$.

For either definition of the structure function, the variability can only be characterized if it is detected above the uncertainties on the measurements.  In practice, this has the following consequences. First, on short time scales where the variability amplitude is small, we are not always able to measure the structure function.  Second, because the two definitions of the structure function are slightly different, the temporal bins where the structure function is undetermined can differ.  This means that for some time intervals, one of the two definitions yields a meaningful measurement of the structure function while the other does not.

For Gaussian distributions of $\Delta m$, $SF$ and $SF_{IQR}$ will be identical, but for distributions with more prominent wings (e.g., an exponential distribution), the value of $SF_{IQR}$ will be smaller than the value of $SF$ corresponding to the same time interval.  $SF_{IQR}$ is particularly robust against outliers that result from bad data and is an especially sensitive measure of variability at short time scales where the amplitude of the variability is small \citep{macleod12}.  Because we are particularly interested in longer time scales with the SBHB candidates, it is not clear that the IQR structure function is preferable, so we consider both forms of the structure function.  Given the distribution of magnitude differences in Figures~\ref{fig:dfdt_c} and \ref{fig:dfdt}, we expect to find $SF_{IQR}(\Delta t)<SF(\Delta t)$.

In calculating the structure function, it is common to include a term to account for the photometric uncertainty.  This can be done by calculating the structure function either of nearby non-varying stars, or \OIII, which is assumed to remain constant.  However, we are assuming here that the \OIII\ flux remains constant and we use it to re-normalize the spectra before calculating $\Delta m$, which, in turn does not allow us to determine the photometric uncertainty in this fashion.  Instead, we propagate the uncertainties in the integrated \OIII\ flux resulting from the spectral decomposition into the error bars on the structure function.  In some cases these uncertainties can be very large, so we calculate the structure function using only measurements with small uncertainties (as determined from three iterations of 3$\sigma$ clipping on the distribution of percent uncertainties in each $\Delta t$ bin). In practice, this approach trims a small number of data points with uncertainties greater than 50\%, leaving the majority of points with uncertainties of less than 10\%. 

\begin{figure}[t]
\centering
\epsscale{2.0} \plottwo{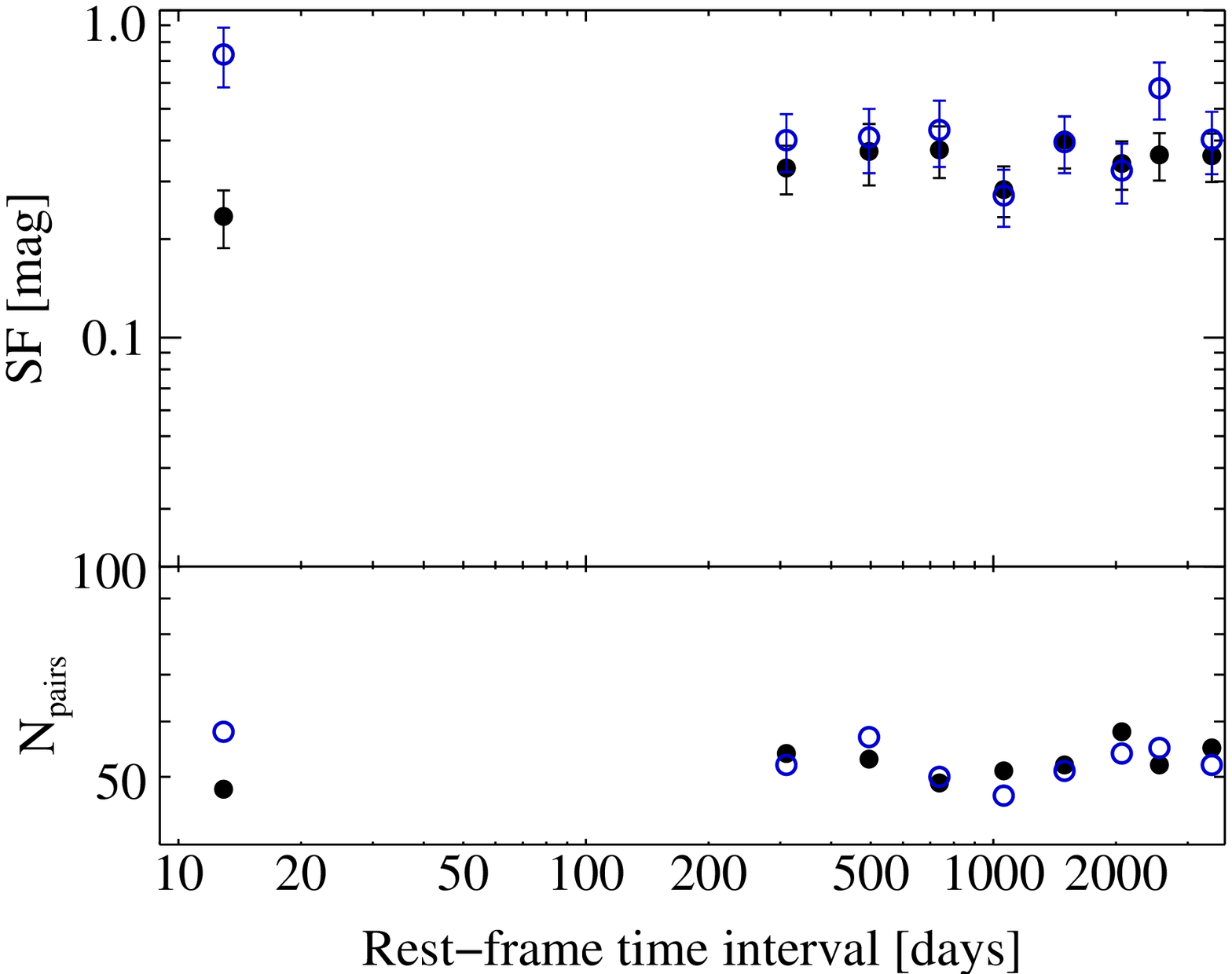}{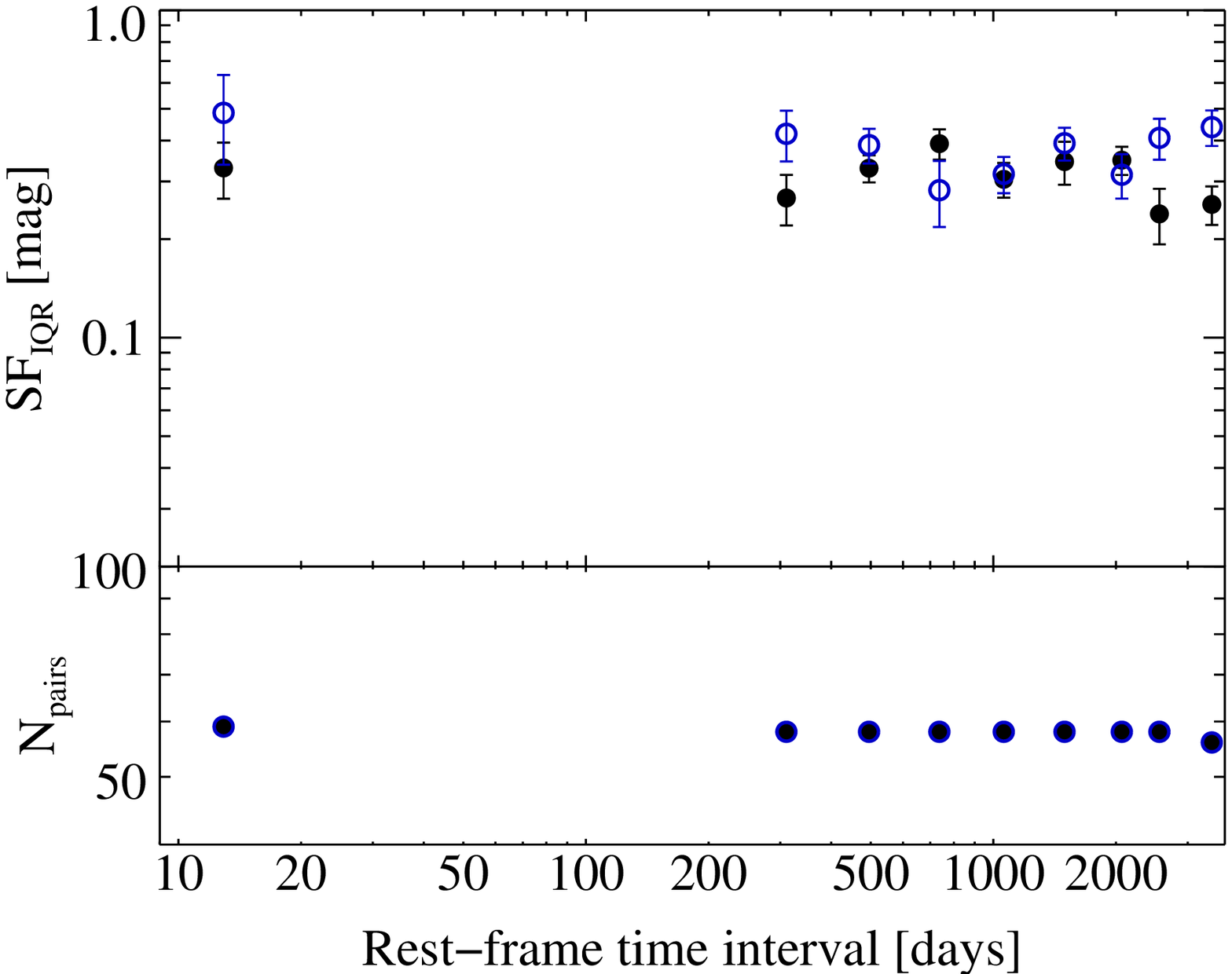} 
\caption{A comparison of  the continuum (black, filled circles) and \Hb\ (blue, open circles) structure functions for the SBHB candidates.  The two versions of the structure function presented in Equations~(\ref{eqn:v04}) and (\ref{eqn:viqr}) are shown in the top and bottom panels, respectively.  At the shortest time scale the value of the structure function here is driven primarily by observations of one object and may not accurately reflect the ensemble variability of the sample.  We find that the shapes of the continuum and emission line structure functions are very similar, i.e. the majority of data points match within the uncertainties.} \label{fig:sf_hb_cont}
\end{figure}

\begin{figure}[t]
\centering
\epsscale{2.0} \plottwo{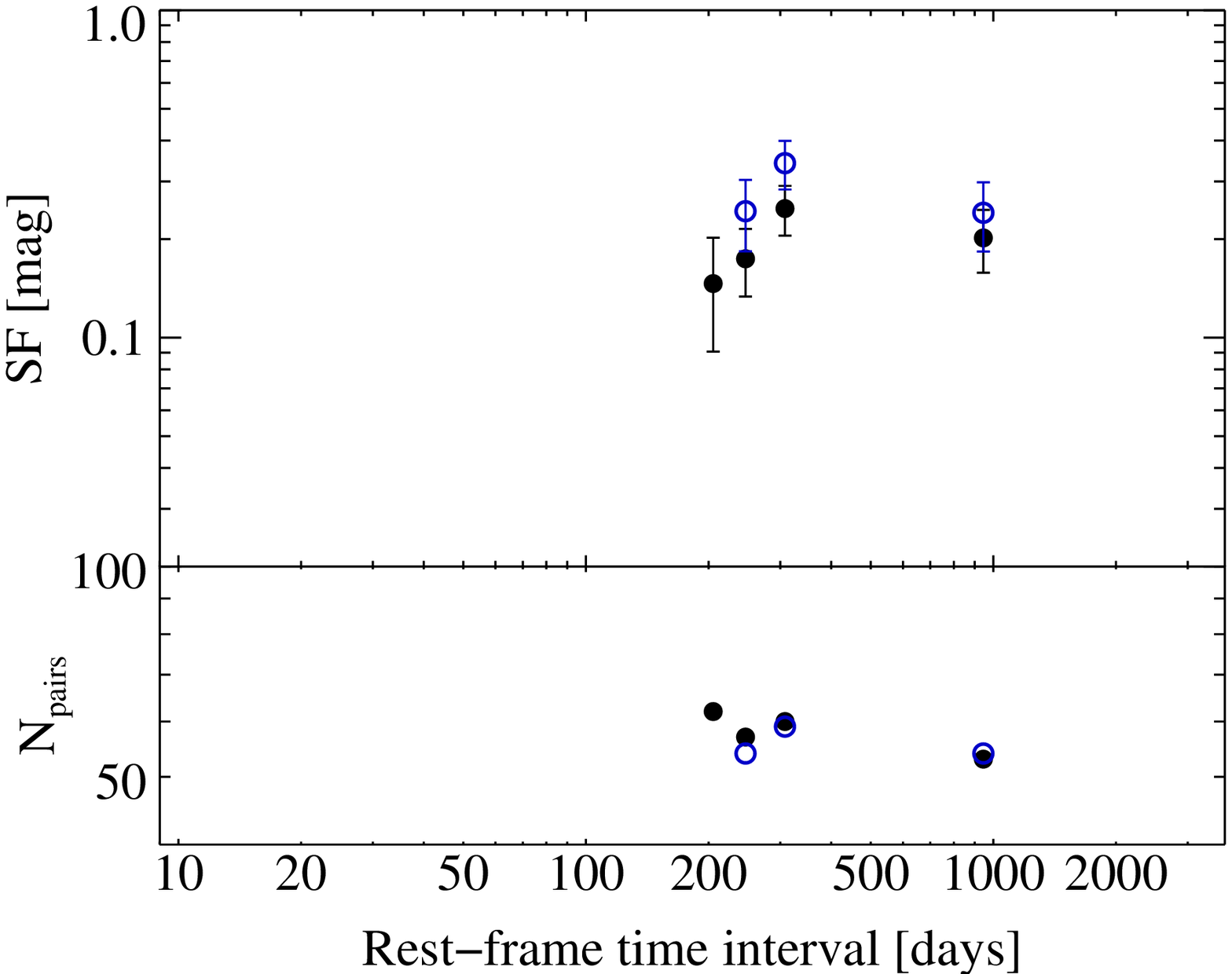}{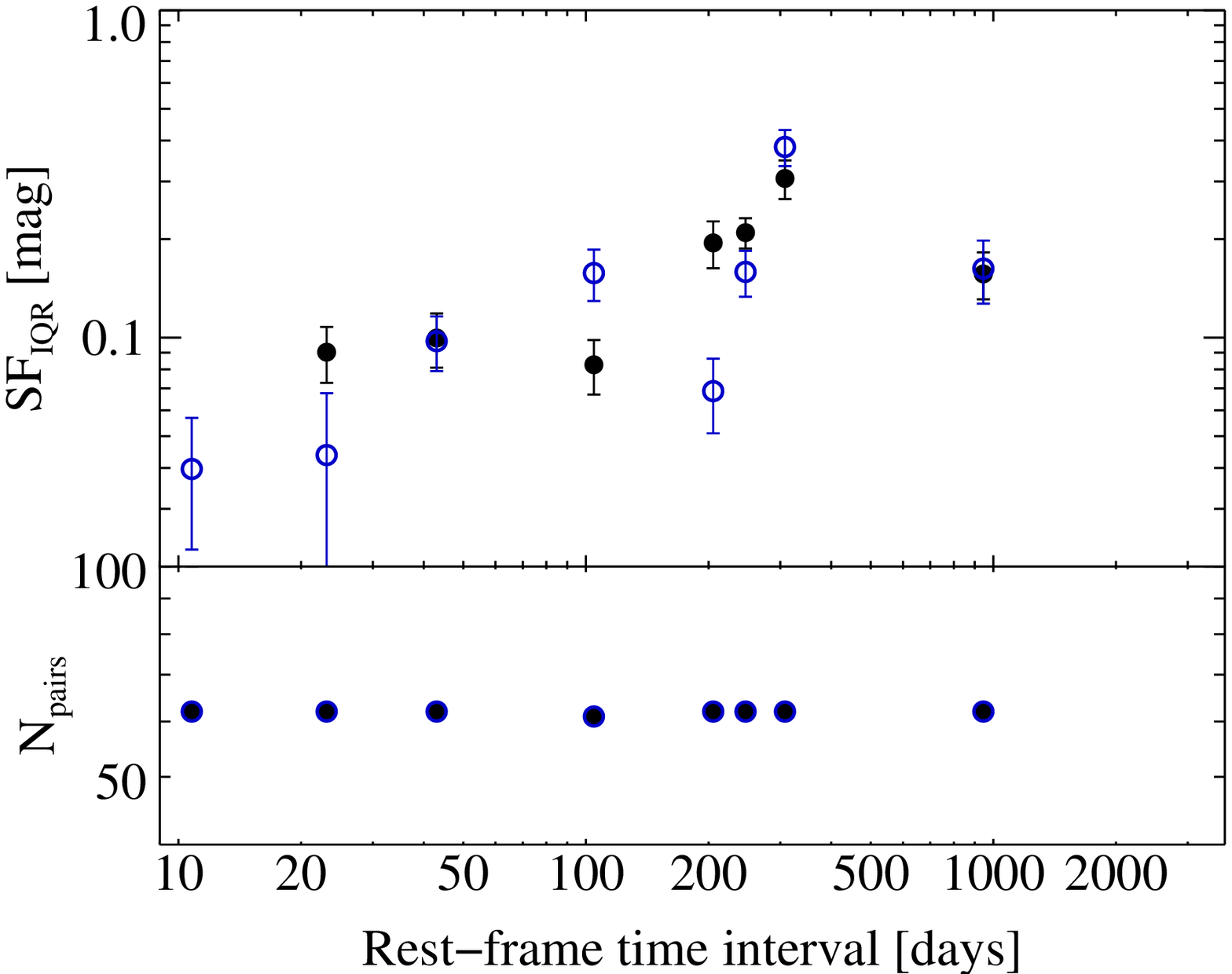} 
\caption{A comparison of  the continuum (black, filled circles) and \Hb\ (blue, open circles) structure functions for the comparison sample.  The \citet{vandenberk04} and IQR structure functions are shown on the top and bottom, respectively.  The shape of the structure function for the continuum and emission line is similar, but not always consistent with each other within the uncertainties.} \label{fig:sf_hb_cont_sdss}
\end{figure}

\begin{figure}[t]
\center\epsscale{2.0} \plottwo{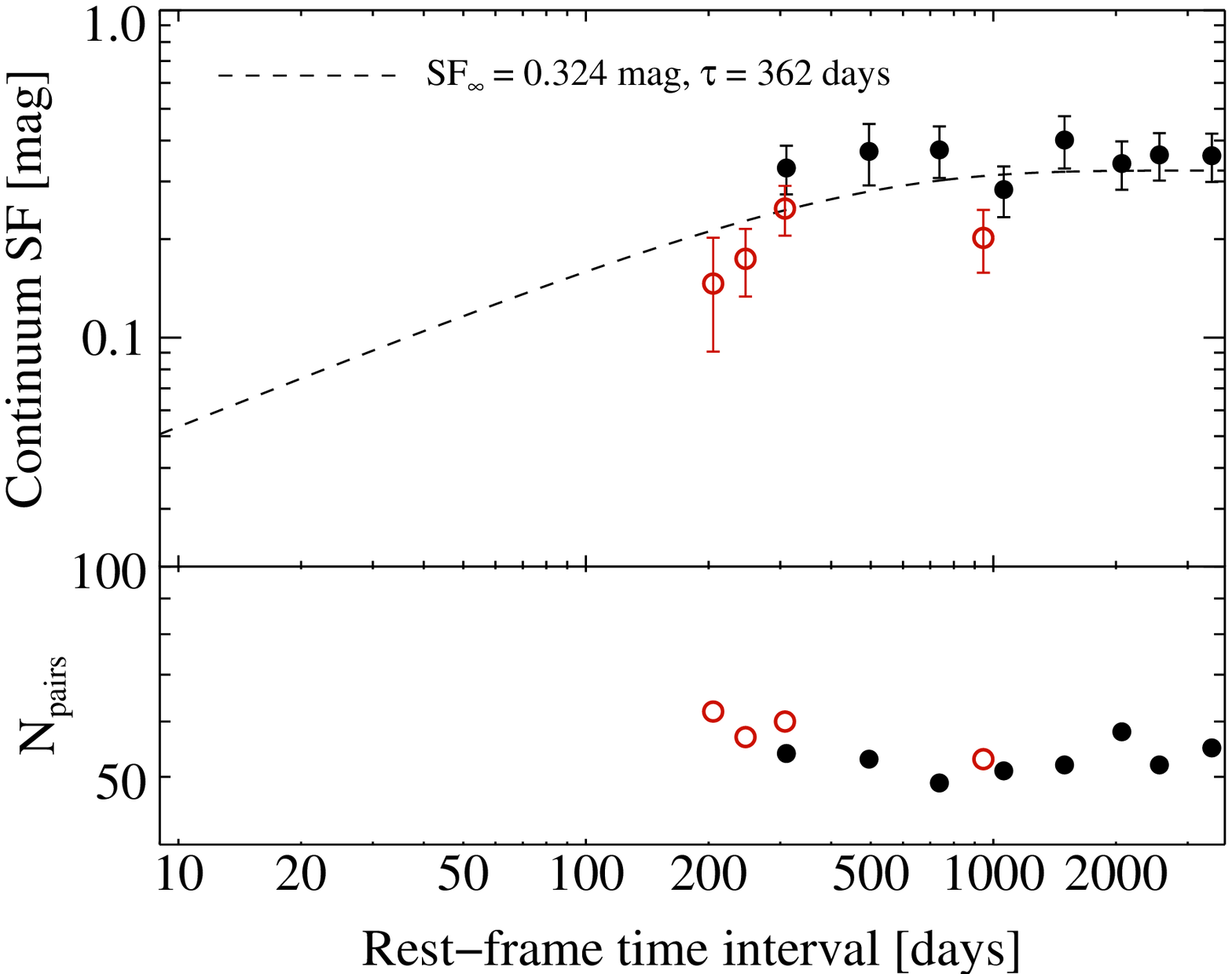}{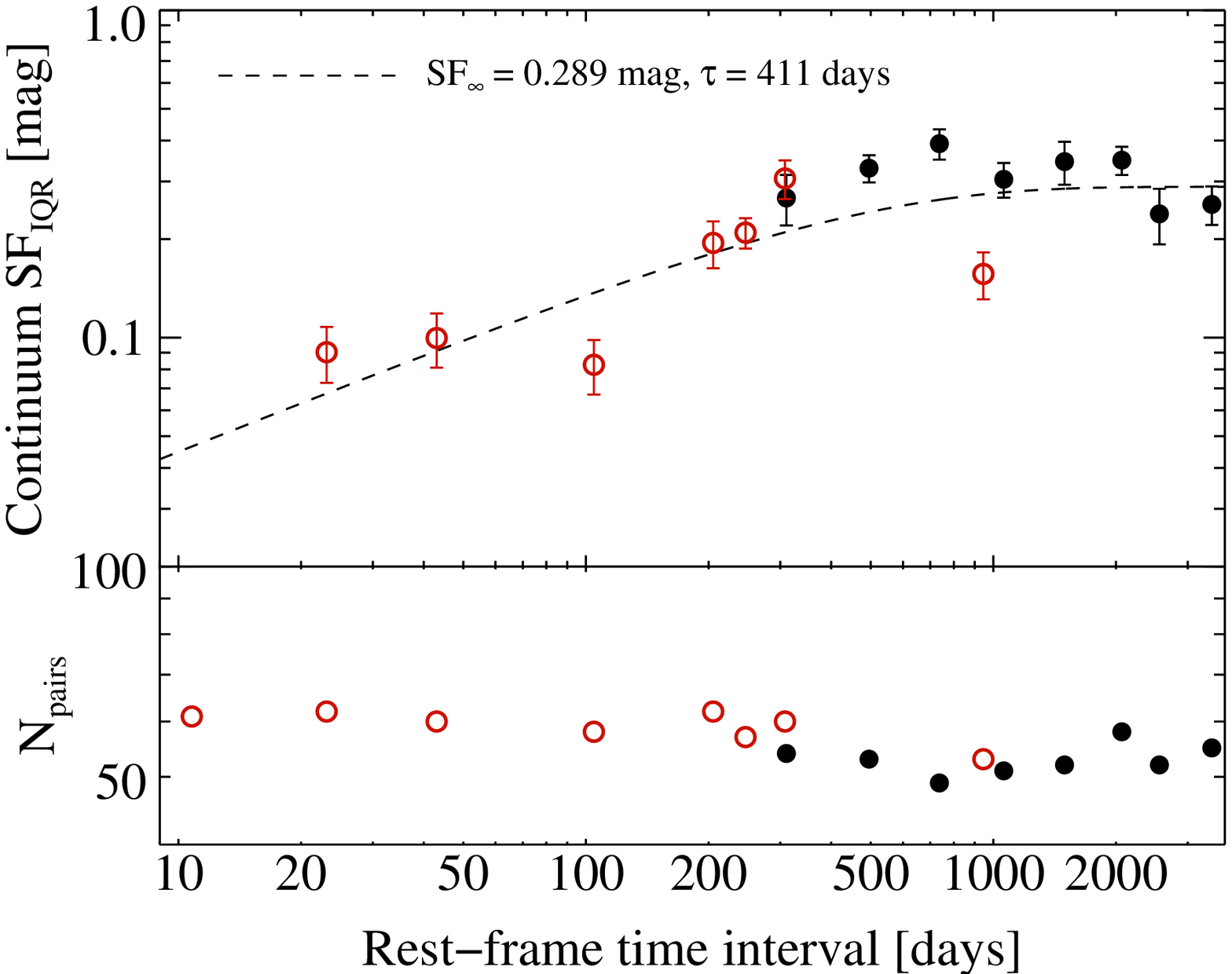}
\caption{A comparison of  the 5100~\AA\ continuum structure functions the SBHB candidates (black, filled circles) and the comparison sample (red, open circles).  The dashed line shows the modified exponential parameterization of combined comparison sample and SBHB structure functions.  See Section~\ref{sec:analysis} for additional details.} \label{fig:sf_cont}
\end{figure}

\begin{figure}[t]
\center\epsscale{2.0} \plottwo{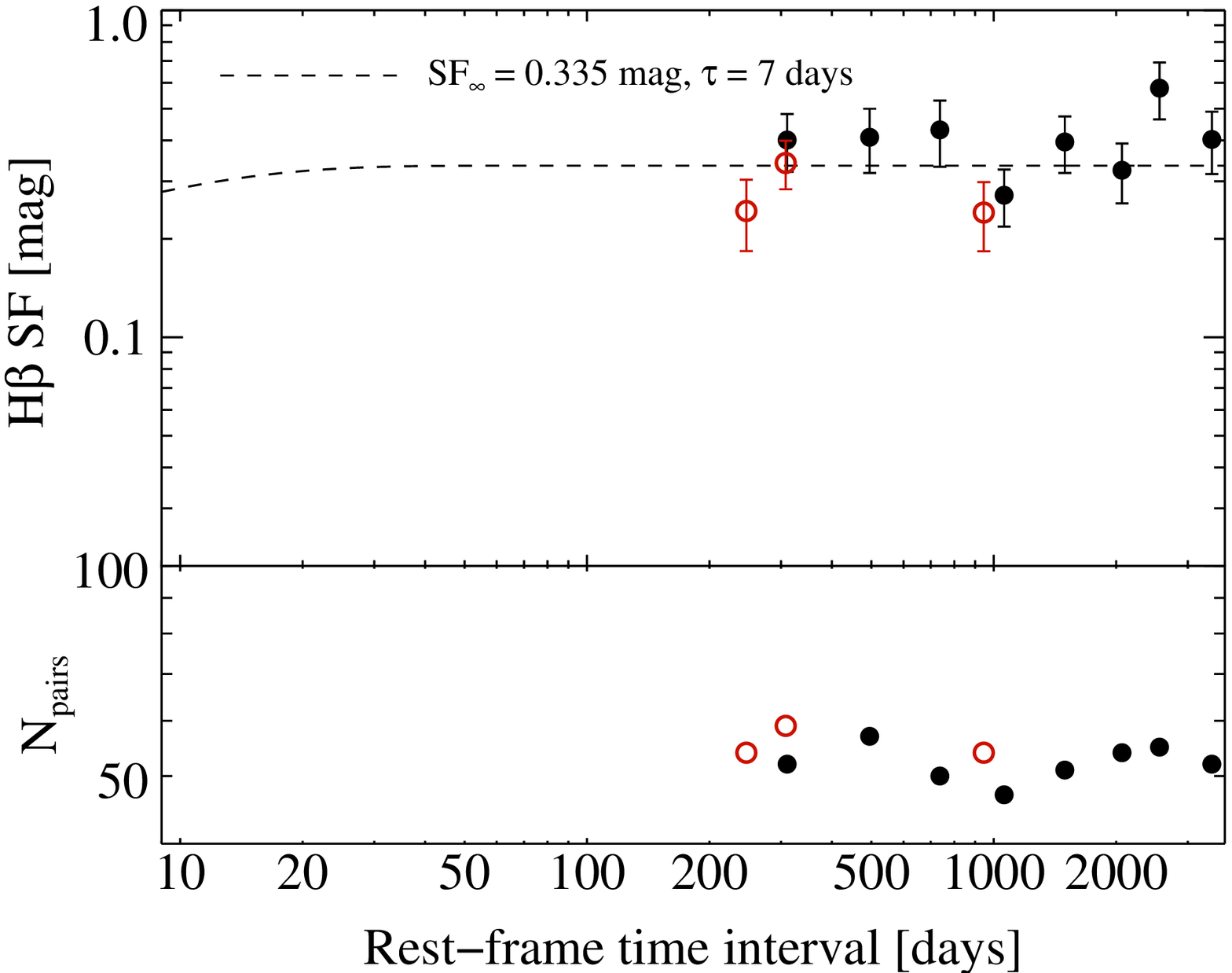}{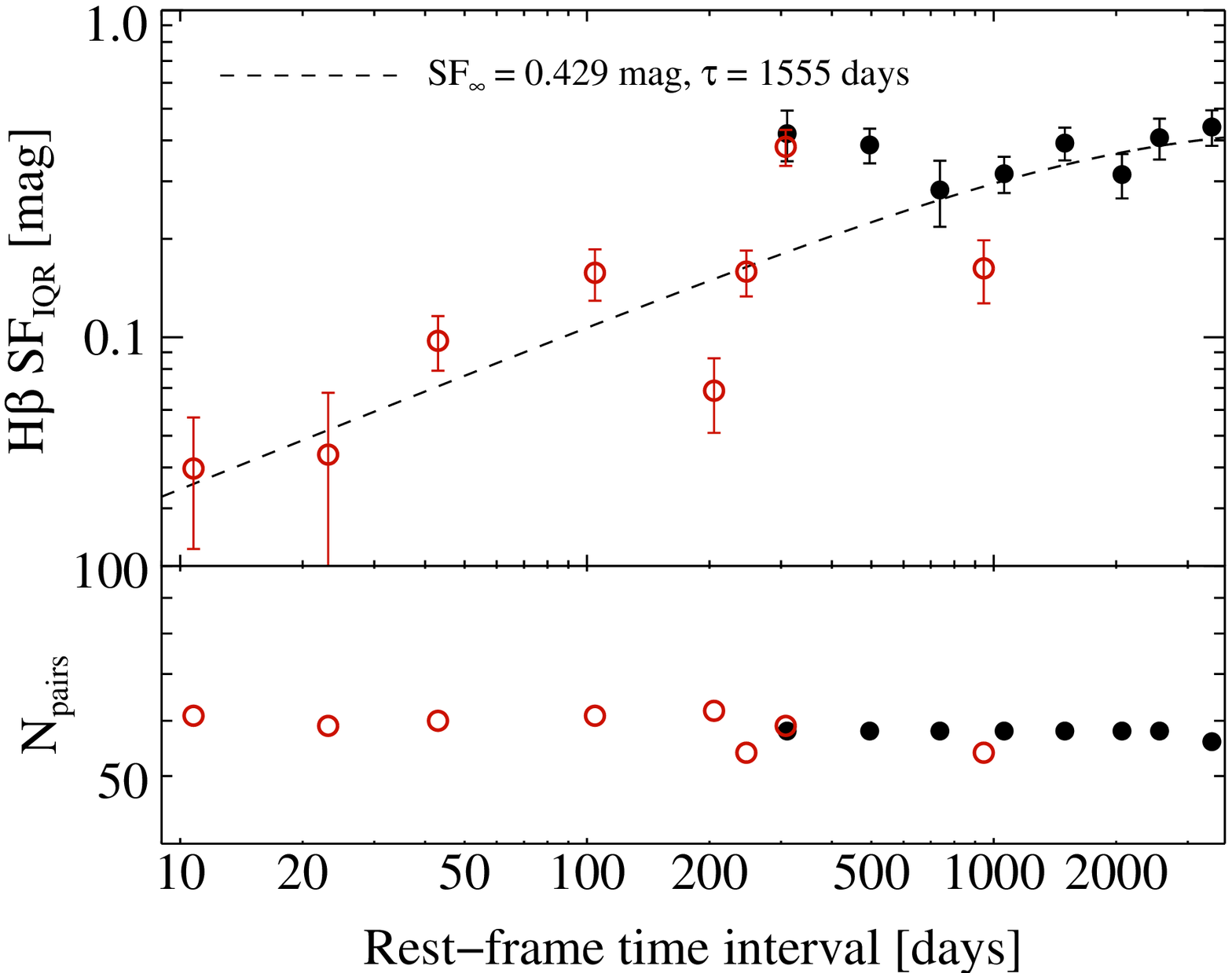}
\caption{A comparison of  the \Hb\ structure functions for the SBHB candidates (black, filled circles) and the comparison sample (red, open circles). The dashed line shows the modified exponential parameterization of combined comparison sample and SBHB structure functions.  See Section~\ref{sec:analysis} for additional details.} \label{fig:sf_hb}
\end{figure}

The continuum and \Hb\ structure functions for the SBHB candidates are shown in Figure~\ref{fig:sf_hb_cont}.  We have selected logarithmic time bins so as to get an equal number of pairs of observations in each bin and adopt this binning for all the structure functions of the SBHB candidates in this work.  We find that the structure functions for the continuum flux density and the integrated \Hb\ line flux agree with each other within the uncertainties for the SBHB candidates.  This is true for both definitions of the structure function that we have adopted here.

For both definitions of the structure function, the variability in the shortest time scale bin ($<20$~days) appears very large.  While most observations in this bin show small variability, the value of the structure function here is driven primarily by many observations of one object which shows large variability.  Thus, we do not regard this point as indicative of the ensemble variability of the SBHB candidates and exclude it from later figures.

\subsection{Comparison to the Sample of Typical SDSS Quasars}
\label{sec:comp}
To assess whether the behavior of the SBHB candidates differs significantly from that of typical quasars, we compare their variability properties to those of the sample of typical quasars presented in Section~\ref{sec:sdss_sample}.  We measure the structure functions of this sample in the same fashion as the SBHB candidates, although we use a new set of temporal bins for this purpose, appropriate for the time sampling of the quasar fluxes.

The magnitude differences of the continuum and the \Hb\ are shown as a function of the rest-frame time interval in Figures~\ref{fig:dfdt_c_sdss} and \ref{fig:dfdt_sdss}, respectively.  The time intervals between observations are clearly concentrated at values below 700~days, much shorter than those available for the SBHB candidates. This is an inevitable consequence of the fact that the spectra of the comparison sample were obtained as part of the SDSS~I--III surveys. In contrast, the follow-up spectra of many of our SBHB candidates were obtained long after their original SDSS spectra, therefore the time interval between those observations are longer. A cursory inspection of these figures shows relatively few objects exhibiting relative flux variations greater than 50\%. The distribution of continuum magnitude differences, shown in the right panel of Figure~\ref{fig:dfdt_c_sdss}, has a median of 0.03 and standard deviation of 0.22. The distribution of \Hb\ magnitude differences, shown in the right panel of Figure~\ref{fig:dfdt_sdss},  has a median of 0.02 a standard deviation of 0.28. We characterize the ensemble variability properties of the comparison sample by calculating the structure function, as discussed in Section~\ref{sec:sf}.  The structure functions for the continuum at 5100~\AA\ and the \Hb\ emission line are presented in Figure~\ref{fig:sf_hb_cont_sdss}.  According to the structure functions, the fractional variability amplitude in this sample is generally on the order of 10--30\%, although it increases with increasing time interval.

In Figure~\ref{fig:sf_cont} we compare the continuum structure function of the SBHB candidates to that of the comparison sample and in Figure~\ref{fig:sf_hb} we show an analogous comparison for the \Hb\ structure functions.  We find that the SBHB candidates display variability that is consistent with that observed for typical quasars in a similar redshift and luminosity range, on the order of 10--30\%.  On the shortest time scales the variability may be even less, but that measurement is hindered by the magnitude of the uncertainties.

To ascertain the robustness of the structure functions and verify the error bars, we simulate the effect on the structure function of perturbing the data within the measured uncertainties.  The $\Delta m$ distribution is redrawn by randomly selecting flux ratios for each object in the sample from a normal distribution with mean and standard deviation equal to the flux ratio measurement and uncertainty, respectively.  The $\Delta m$ distribution is then propagated into a structure function assuming the observed uncertainties.  We repeat this process $10^3$ times and determine the median of the $SF$ and $SF_{IQR}$ distributions in each time bin and their associated standard deviations.  We find that this method reproduces the observed flux ratio distribution within the uncertainties and preserves the mean. We conclude that our uncertainties are reasonably characterized and that the match between the structure functions of the SBHB candidates and the comparison sample is robust.

We also ask what the structure functions would look like with zero variability, given the observed uncertainties.  This is simulated in an identical fashion, except the flux ratios are generated from a normal distribution around unity.  We find that zero variability gives a flat structure function near but not equal to zero.  This occurs because the structure function involves the root-mean-square of the absolute values of magnitude differences, therefore, positive and negative excursions from the mean magnitude do not cancel out. This confirms that, within the observed errors, the observed structure functions are not consistent with zero variability. 

In view of the different distributions of time intervals for the SBHB candidate sample and the comparison sample, we can only make a direct comparison of the measured structure functions for time intervals between 300 and 900 days. To assess the similarities between the structure functions of the two samples further, we ask whether the models in the literature that are used to describe the structure functions of typical quasars can also describe the structure functions of the SBHB sample.

The structure functions of typical quasars are generally well parameterized by simple power-law functions \citep[e.g.,][]{vandenberk04,macleod12} as well as modified exponential functions that are motivated by damped random walk models.  Because we see evidence of a turn-over in the structure functions of the SBHB candidates on time scales around 1000~days, we adopt a modified exponential here.  There is some variation in the exact form that is adopted, for example \citet{vandenberk04} adopt a power-law model, but we choose the form used by \citet{macleod12} because it is able to reproduce the break that we see in our structure functions.  The parameterization is
\begin{equation}
\label{eqn:modexp}
SF = SF_{\infty} \left( 1- e^{-\Delta t/\tau}\right)^{1/2},
\end{equation}
where $SF_{\infty}$ and $\tau$ are adjustable parameters describing the normalization and break time, respectively.  The resulting fits are also shown in Figures~\ref{fig:sf_cont} and \ref{fig:sf_hb} with the best-fit parameters listed in Table~\ref{tab:sf_fits}.  Note that we find $\tau \sim 10$~days for the \Hb\ structure function calculated from Equation~\ref{eqn:v04} (shown in the top panel of Figure~\ref{fig:sf_hb}).  This value is not constrained because the variability on short time scales cannot be distinguished above the noise, hence no turnover is observed in the structure function.

\begin{deluxetable}{lccc}
\tabletypesize{\small}
\tablecolumns{5}
\tablewidth{0pc}
\tablecaption{Structure Function Parameterization\tablenotemark{a}
\label{tab:sf_fits}
}
\tablehead{
\colhead{Type} & 
\colhead{$SF_{\infty}$} & 
\colhead{$\tau$} & 
\colhead{$\chi^2/\nu$} \\ 
\colhead{} & 
\colhead{(mag)} & 
\colhead{(days)} & 
\colhead{} \\ 
\colhead{(1)} & 
\colhead{(2)} & 
\colhead{(3)} & 
\colhead{(4)} 
}
\startdata
Continuum
 & 
0.324$\pm$0.026
 & 
\phantom{0}360$\pm$150
 & 
2.8
\\
Continuum IQR
 & 
0.289$\pm$0.017
 & 
\phantom{0}410$\pm$90
 & 
5.9
\\
\Hb
 & 
0.335$\pm$0.021
 & 
\phantom{0}8$\pm$6\tablenotemark{b}
 & 
2.1
\\
\Hb\ IQR
 & 
0.429$\pm$0.058
 & 
1560$\pm$590
 & 
8.1
\\
\enddata
\tablenotetext{a}{The modified exponential function is defined in Equation~\ref{eqn:modexp}.  All parameterizations listed here are for the combined binary candidate and SDSS comparison samples.}
\tablenotetext{b}{This value is not constrained, as discussed in Section~\ref{sec:comp}.}
\end{deluxetable}


We find that the parameterizations of the combined SBHB candidate and comparison structure functions here are generally consistent with those in the literature.  For structure functions of the continuum light between $4000$ and $6000$~\AA, sampled on rest-frame time scales of approximately 1-1000~days, \citet{macleod12} find that their data are well fit by  $SF_{\infty}=0.222\pm0.13$~mag and $\tau=638\pm98$~days.  This break and normalization are comparable to the parameterizations for both continuum structure functions.

In the range of time intervals from 300 to 900 days, the structure functions of the SBHB candidates and the comparison sample overlap and are in good agreement with each other. Furthermore, the structure functions of the SBHB candidates and comparison sample can be well fit with a common model described by parameters comparable to those obtained for samples of normal quasars.  This suggests that the SBHB candidates exhibit variability consistent with that expected for regular quasars on timescales $>900$ days.  Hence, we conclude that the variability properties of the binary properties are similar to those of regular quasars.

\section{Implications of Observed Variability Properties}
\label{sec:implications}

Our main, new observational result is that, on time scales longer than approximately one year the variability of both the continuum flux and the broad \Hb\ emission line flux in the SBHB candidates is consistent with the 10--30\% observed in typical quasars of similar redshift and luminosity.  Additionally, the continuum structure functions of the SBHB candidates and the quasars of the comparison sample can be described by a common model. The same is true of the \Hb\ structure functions. Since we are only able to sample the stochastic variability of the SBHB candidates in a statistical sense and on relatively short time scales, we cannot carry out direct tests of specific models for the interaction of a SBHB with a circumbinary disk. The models of \citet{dorazio13} and \citet{farris14} predict periodic variability on time scales ranging from about half to about 10 times the orbital period. The variability arises from accretion rate fluctuations on the orbital period which can also lead to flux variability because of beaming with half that period. On time scales longer than the orbital period, the variability is caused by tidal interactions of the binary with the circumbinary disk, leading to bright spots in the disk. These time scales would amount to several decades to many centuries for the SBHB candidates in our sample, in sharp contrast to the time scale we have probed, which are $\sim 1\;$month to $\sim 12\;$years. In order to carry out direct tests of the models, we would need to obtain light curves of individual objects spanning much longer temporal baselines.

These results suggest three possible conclusions.  (a) The binary candidates are, in fact, not SBHBs, and the mechanism responsible for the offset broad lines does not leave an imprint on the variability properties of these objects.  In this case, the binary candidates still represent a very interesting population of quasars where some mechanism in the BLR has produced offset broad emission lines, as well as the correlation noted in \pone\ and reproduced here where the lines are skewed in the direction of their shift. For example, such a mechanism may be a 1-arm spiral or a warp that leads to a high-contrast, non-axisymmetric perturbation in a disk-like BLR \citep[see examples in][respectively]{lewis10,wu08}. Thus, the peak of the broad line is displaced according to the projected velocity of the brightest portions of the disk. (b) Some objects in the sample are SBHBs, but the contamination of our sample by objects that are not is enough to prevent our detecting a difference, even though SBHBs have different variability properties.  (c) SBHBs have the same variability properties as regular quasars.  In order to say more about the first two points, we require the accumulation of additional lines of evidence, particularly the radial velocity curves that will be the focus of an upcoming paper.

In the context of the third possibility, our main result suggests that, if the observed \Hb\ flux variability is the result of reverberation of the continuum, then the extent of the BLR of the SBHB candidates is comparable to that of typical quasars. Thus, in the remainder of this section, we consider the implications of these results for the structure of the BLR in the SBHB candidates. We restrict our discussion to the specific scenario in which the offset broad emission lines arise in a BLR that is associated with the secondary BH in a SBHB.  \citet{shen13a} and \citet{ju13} have already attempted to constrain the orbital properties of the SBHB population based on the observed properties of the candidates.  Here, we are interested in examining a more specific question in the context of our variability data: if the BLRs of SBHB candidates have a similar extent to those of typical quasars, what are the ensuing constraints on the orbital parameters of the hypothesized SBHBs? To this end, we compare the extent of a typical BLR to characteristic length scales in the binary, the size of the Roche lobe of the accreting black hole and the radius at the which the accretion disk of the accreting black hole is truncated because of the tidal effect of its companion.

We begin by collecting here a number of constraints derived by combining observations and basic orbital mechanics.  As in \pone, we assume that only the secondary (less massive) BH is active, in other words, all observed properties of the AGN, such as the bolometric luminosity and the Eddington ratio, refer to the secondary. Adapting equations  (1) and (2) of \pone, we write the orbital period and separation of the binary as
\begin{equation}
P = {2625\; M_8\over (1+q)^3\, V_{2,3}^3}\;{\rm yr}
\label{eqn:period}
\end{equation}
and
\begin{equation}
a = {0.432\; M_8\over (1+q)^2\, V_{2,3}^2}\;{\rm pc,}
\label{eqn:separation}
\end{equation}
where $q = M_2/M_1<1$ is the mass ratio, $M_8$ is the {\it total} mass of the binary in units of $10^8\;{\rm M}_\odot$, and $V_{2,3}$ is the true orbital velocity of the secondary in units of $10^3$~km~s$^{-1}$, assuming a circular orbit. In \pone\ we presented the observed distribution of {\it projected} orbital velocities, $u_2=V_2\,\sin i\sin\phi$ (where $i$ and $\phi$ are the inclination and phase angles of the binary), spanning a range from one to a few thousand km~s$^{-1}$. Therefore, as long as  $i$ and $\phi$ do not have extreme values, we expect $V_{2,3}$ to be of order a few. Combining equations (\ref{eqn:period}) and (\ref{eqn:separation}) we can cast Kepler's third law for such a binary as
\begin{equation}
  a = 0.228\; M_8^{1/3}\,\left(P\over 10^3\;{\rm yr}\right)^{2/3}\;{\rm pc}\; .
\label{eqn:Kepler}
\end{equation}
  
\begin{figure}[t]
\centerline{\epsscale{1.2} \plotone{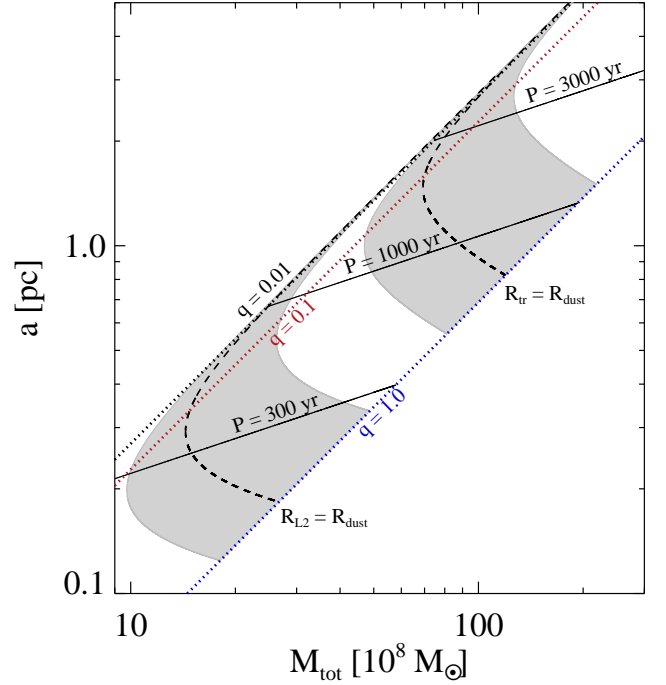} }
\caption{Constraints on the binary parameters of the SMBH candidates derived by considering the size of the BLR in comparison to other characteristic lengths in the binary. The dotted lines are lines of constant mass ratio for a fixed orbital velocity of the secondary BH. The thin solid lines are lines of constant orbital period. The conditions that the BLR is not tidally truncated and is smaller than the Roche lobe are depicted by grey bands. More details can be found in Section~\ref{sec:implications} of the text.} \label{fig:constraints}
\end{figure}

In Figure~\ref{fig:constraints} we show graphically the physical parameter space described by the binary separation, $a$ (in pc), and the total mass of the two BHs, $M_8(\equiv M_{tot}/10^8\;{\rm M}_\odot)$.  The dotted lines in Figure~\ref{fig:constraints} show the relation between $a$ and $M_8$ ($a\propto M_8$; see eqn.~\ref{eqn:separation}), for a fixed value of the orbital speed of the secondary BH ($V_2=4000\;{\rm km\;s}^{-1}$), and three different values of the mass ratio ($q=0.01$, 0.1, and 1). These lines effectively delineate the entire range of combinations of $a$ and $M_8$ since the $q=0.01$ line is nearly identical to the $q=0$ line and since $q>1$ contradicts our basic assumption that only the secondary BH is active.  The solid lines in Figure~\ref{fig:constraints} are lines of constant orbital period, following Kepler's third law ($a\propto M_8^{1/3}$; see eqn~\ref{eqn:Kepler}). If the orbital speed of the secondary BH increases, then the lines of constant $q$ move towards the lower right; their horizontal shift corresponds to a change in $M_8$ by a factor $(V_2/4000\;{\rm km\;s}^{-1})^2$ as prescribed by equation~(\ref{eqn:separation}).

\begin{figure}[t]
\begin{center}
\epsscale{1.2} \plotone{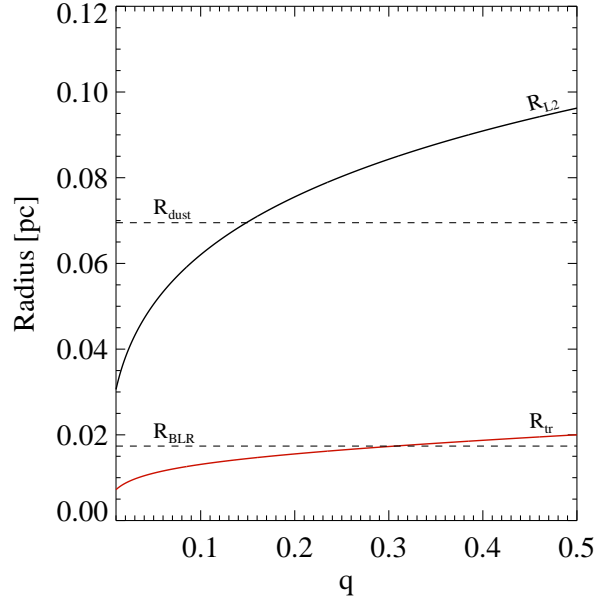} 
\caption{Comparison of the secondary Roche lobe radius ($R_{L2}$, solid black line) and truncation radius ($R_{tr}$, solid red line) to the empirically calibrated BLR radius ($R_{BLR}$, lower dashed line) and dust sublimation radius ($R_{dust}$, upper dashed line). In this particular realization we have assumed a binary with a total mass of $10^8\;{\rm M_\odot}$ ($M_8=1$) and an orbital separation of $a=0.3$~pc and we plot $R_{L2}$ and $R_{tr}$ as a function of the mass ratio, $q$. We take $R_{BLR}=0.017$~pc and $R_{dust}=0.068$~pc; these are fixed since they depend on the luminosity of the active BH. More details can be found in Section~\ref{sec:implications} of the text.} \label{fig:binradii}
\end{center}
\end{figure}

In order for gas in the vicinity of the secondary black hole to be gravitationally bound to it, the gas must be within its Roche lobe.  The following parametric expression for the effective radius of the Roche lobe of the secondary, $R_{L2}$ (the radius of a sphere with the same volume as the Roche lobe), relative to the binary separation, $a$, is given by \citet{eggleton83}
\begin{equation}
  {R_{L2}\over a}\equiv x_{L}(q)\approx 
  \frac{0.49\,q^{2/3}}{0.6\,q^{2/3}+\ln(1+q^{1/3})}.
\label{eqn:xlobe}
\end{equation}
At some critical distance from the secondary, the accreting object stops dominating the potential, which departs significantly from spherical symmetry with the result that orbits (coplanar with the binary orbit) at that radius cannot be circular. Therefore an accretion disk around the secondary is truncated at this critical distance because particle orbits start to cross each other. Detailed calculations of the disk truncation radius are presented by \citet[][and references therein]{paczynski77}. For our purposes we adopt the following approximate expression for the truncation radius relative to the Roche lobe radius by \citet[][section 6.1, eqn 6.8]{eggleton11}
$$
\frac{R_{tr}}{R_{L2}} \sim \left(1.9+\frac{2.2}{q}\right)\,x_{L}^3(q),
\qquad\qquad\qquad
$$
\begin{equation}
\qquad\qquad
\quad\Rightarrow\quad 
\frac{R_{tr}}{a} \sim \left(1.9+\frac{2.2}{q}\right)\,x_{L}^4(q), 
\label{eqn:Rtrunc}
\end{equation}
where $x_{L}(q)$ is given by equation (\ref{eqn:xlobe}). In Figure~\ref{fig:binradii}, we illustrate how the secondary Roche lobe radius and the truncation radius vary with the mass ratio for a binary with $a=0.3$~pc and $M_8=1$ (hence $P\approx 1500$~yr). For a larger value of $a$, $R_{L2}$ and $R_{tr}$ increase proportionally. However, $R_{tr}$ is always small compared to $R_{L2}$ and $a$. 

To estimate the extent of the BLR we start with the empirical radius-luminosity ($R_{BLR}$\,--$\,L$) relationship of  \citet{bentz13}, connecting the reverberation lag of the broad H$\beta$ flux, $\tau_{\rm H\beta}$, to the monochromatic luminosity of the AGN at 5100~\AA.  This relationship is calibrated over four orders of magnitude in monochromatic luminosity and has some scatter, whereas our sample spans a relatively narrow luminosity range (about an order of magnitude). To account for the scatter, we do not apply the $R_{BLR}$\,--$\,L$ directly but instead we consider the measured lags of the AGNs in \citet{bentz13} with luminosities in the same range as the AGNs of our sample. For this narrow range of luminosities there is no discernible correlation therefore we simply adopt the median lag to obtain an estimate of $R_{BLR} = c\,\tau_{\rm H\beta} = 0.017^{+0.014}_{-0.006}\;$pc, corresponding to a monochromatic luminosity of $\log[\lambda L_\lambda(5100\;{\rm A})/{\rm erg\;s}^{-1}] = 43.62$. The upper and lower bounds on $R_{BLR}$ define the intervals within which we find $1/3$ of the \citet{bentz13} AGNs above and below the best value of $R_{BLR}$, respectively (analogous to ``$\pm 1\sigma$'' intervals). The extent of the BLR (its effective outer radius) is likely a few times larger than $R_{BLR}$, as evidenced by the following observational results: (a) The H$\alpha$ line responds to continuum variations with a lag that is approximately twice as long as that of H$\beta$ \citep[e.g.,][]{grier13,bentz10b}. Similarly, the lines of the optical \ion{Fe}{2} complex respond to continuum variations with a lag that is approximately 2--3 times as long as that of H$\beta$ \citep{barth13}. (b) The dust sublimation radius, as inferred from interferometric imaging and from the reverberation lag of the near-IR continuum relative to the optical continuum is 4--5 times larger than $R_{BLR}$ \citep[e.g.,][]{kishimoto11, suganuma06, koshida14}. In view of the above results, we take $R_{dust} = 4\,R_{BLR}$ to be a plausible estimate of the maximum extent of the line-emitting gas in the BLR of a typical quasar. Because of the scatter in the $R_{BLR}\,$--$\,L$ relation we illustrate the outer extent of the BLR as a broad band rather than a sharp line in Figure~\ref{fig:constraints}. In the illustration of Figure~\ref{fig:binradii} we show the BLR and dust sublimation radii as dashed lines. For the combination of binary parameters adopted in that illustration, $R_{BLR}$ can be either smaller or larger than $R_{tr}$ depending on the value of $q$, but it is always smaller than $R_{L2}$. In contrast, $R_{dust}$ is always larger than $R_{tr}$, but it can be either smaller or larger than $R_{L2}$ depending on the value of $q$.

If we suppose that the extent of the BLR of SBHB candidates is the same as that of typical quasars, then we obtain lower limits on the orbital separation and total mass in the binary.  We consider two scenarios to place constraints on the binary properties, amounting to two ways of placing the truncation radius relative to the dust sublimation and BLR radii (see example in Figure~\ref{fig:binradii}).

In the first, more extreme scenario, we require that $R_{dust}\lsim R_{tr}$, which is equivalent to assuming the BLR has the same extent as in a typical quasar of a given luminosity.  This condition can also be viewed as a requirement that the potential in the outskirts of the BLR is still dominated by the secondary BH.  If we adopt this scenario, we effectively assume that the binary separation is large enough that the BLR must not be tidally truncated by the primary black hole.  Thus we obtain the dashed black line at the upper right corner of Figure~\ref{fig:constraints}, which is based on the best estimate of $R_{dust}$. The grey band represents the uncertainty in $R_{dust}$ about the best estimate, arising from the scatter in the correlation between $R_{dust}$ and quasar luminosity. The allowed values of $a$ and $M_8$ are in the strip between the extreme values of $q$ and above the $R_{tr}=R_{dust}$ band.  This leads to larger orbital separations and we infer large total masses, $\gsim 10^{10}\;{\rm M}_\odot$, and very long orbital periods, $\gsim 10^{3}\;$yr.

In the second scenario we require that $R_{dust}\lsim R_{L2}$, which is equivalent to assuming only that the BLR is bound to the secondary BH.  This case is more conservative in the sense that it results in less stringent constrains on the properties of the binary given that the extent of the BLR is set by the observed quasar luminosities and $R_{L2} > R_{tr}$.  This more generous assumption leads to masses $\gsim 10^{9}\;{\rm M}_\odot$ and periods $\gsim 10^{2}\;$yr, as indicated by the dashed black line and grey band at the lower left corner of Figure~\ref{fig:constraints}. We consider this possibility more likely because the emission line profiles of our SBHB candidates are generally asymmetric, unlike the typical quasar population (see, for example, Figure~5 of \pone). Their asymmetries suggest that the structure of the BLR in SBHB candidates is not typical; this may result from the tidal effect of the primary BH onto gas that is bound to the secondary BH. The above constraints are modified if we adopt a higher value of the orbital speed of the accreting BH. Specifically, we can set $V_2=8000\;{\rm km\;s}^{-1}$ without violating the observational constraints, which would lead to an allowed mass range that is four times higher than that depicted in Figure~\ref{fig:constraints} for a given orbital separation, and orbital periods that are approximately a factor of two shorter. 

The question of the allowed orbital periods is important because it has a bearing on the detectability of radial velocity variations. If the expected orbital periods are of order $10^3\;$yr, as in the first scenario described above, then we would not be able to detect significant radial velocity variations in our ongoing spectroscopic monitoring program with observations spanning a decade \citep[see illustration in Fig.~5 of][]{decarli13}. Thus, if we do detect significant velocity variations, we may conclude that, {\it in the context of the SBHB scenario} one or more of the following may be true: (a) the mass of SBHB is $\gsim {\rm several}\times 10^{9}\;{\rm M}_\odot$, as long as the secondary BH is the one that is active, (b) the BLR is not confined within the truncation radius defined by equation~(\ref{eqn:Rtrunc}), (c) the BLR is associated with the primary BH. Possibility (c) implies that $q>1$, effectively, which corresponds to the lower right corner of Figure~\ref{fig:constraints} where the equations above no longer apply and must be re-cast accordingly. 

\section{Summary}
\label{sec:summary}
We have presented new spectroscopic observations of a sample of 88 quasars that are candidate SBHBs. These were selected in \pone\ based on the fact that their broad \Hb\ emission lines are displaced from the frame of their host galaxies by $\sim 1000$--5000~km~s$^{-1}$, which was taken as the signature of motion of the accreting BH around an unseen companion. Moreover, in the first round of followup observations, presented in \pone, $\sim20$\% of the objects observed showed changes in the velocity offsets of their broad \Hb\ lines that were consistent with binary motion.  Our primary goal in this paper was to present the new observations and improvements in our analysis methodology, including a more extensive analysis of the uncertainties. The new observations extend the temporal baseline of our ongoing monitoring campaign and form the basis for a study of radial velocity variations, which will be the subject of a forthcoming paper.

We have also taken this opportunity to study the variability of the optical continuum and the integrated flux of the broad \Hb\ emission line in the sample of SBHB candidates. We compared the variability properties of this sample with those of a comparison sample of typical quasars of similar redshift and luminosity with multiple epochs of spectroscopy available from the SDSS.  We decomposed the optical spectra to separate the quasar continuum, broad \Hb, and narrow \OIII\ emission and calculated spectral properties.  These measurements, which are of general interest for the sample but are not all used in this work, are presented in an appendix along with a careful analysis of the associated uncertainties.

For each sample, we calculated the structure function for the optical continuum and broad \Hb\ emission line.  We found that the ensemble variability of the SBHB candidates is consistent with that of the comparison sample of typical quasars on time scales of approximately one year or longer. Moreover, the continuum and \Hb\ structure functions of the two samples are described by a common model, with parameters applicable to typical quasars.  This result is robust in the sense that it cannot be a consequence of measurement uncertainties and is inconsistent with no variability.  Furthermore, the amplitude of variability of the broad \Hb\ emission line on a given time scale is comparable to that of the optical continuum measured at 5100~\AA. One possible conclusion is that, taken together and assuming that the \Hb\ flux variability is driven by the continuum variability, these results suggest that the extent of the BLR is SBHB candidates is similar to that in typical quasars. But other interpretations are possible, as we note at the beginning of section~\ref{sec:implications}.

We have considered the implications of these results in the context of the binary hypothesis. If we suppose that the BLR of SBHB candidates is similar to that of typical quasars and is not truncated by the binary companion (our first scenario), we infer long orbital periods ($\gsim 10^3\;$yr) and large total masses ($\gsim 10^{10}\;{\rm M}_\odot$). If, however, we only require that the BLR gas is contained within the Roche lobe of the accreting BH and its outer parts can feel the tidal influence of the companion (our second scenario, a more likely possibility in view of the skewed and asymmetric \Hb\ profiles), we infer orbital periods and masses that are approximately an order of magnitude lower. These inferences can be tested by measurements or limits on radial velocity variations, which will be the focus of a forthcoming paper.

\acknowledgements

This work was supported by grant AST-1211756 from the National Science Foundation. We thank the anonymous referee for comments and suggestions. JR acknowledges helpful discussions with Jonathan Trump, Kate Grier, and Mike DiPompeo.  ME thanks the members of the Center for Relativistic Astrophysics at Georgia Tech and the Department of Astronomy at the University of Washington, where he was based during  the early stages of this work, for their warm hospitality. TB acknowledges support from the Alfred P. Sloan Foundation under Grant No. BR2013-016 and support from the National Aeronautics and Space Administration under Grant No. NNX15AK84G issued through the Astrophysics Theory Program.  We thank Yue Shen for his help with the SDSS comparison sample and Stephanie Brown for her help in proofreading the manuscript. We also thank the staff at Kitt Peak National Observatory, Apache Point Observatory, and the Hobby-Eberly Telescope for their expert help in carrying out the observations.  

This work is based on observations obtained with the Apache Point Observatory 3.5-meter telescope, which is owned and operated by the Astrophysical Research Consortium. 

This work is based on observations at Kitt Peak National Observatory, National Optical Astronomy Observatory (NOAO Prop. ID: 2014A-0098; PI: Runnoe), which is operated by the Association of Universities for Research in Astronomy (AURA) under cooperative agreement with the National Science Foundation.

The Hobby-Eberly Telescope (HET) is a joint project of the University of Texas at Austin, the Pennsylvania State University, Stanford University, Ludwig-`Maximillians-Universit\"at M\"unchen, and Georg-August-Universit\"at G\"ottingen. The HET is named in honor of its principal benefactors, William P. Hobby and Robert E. Eberly.

The Marcario Low-Resolution Spectrograph is named for Mike Marcario of High Lonesome Optics, who fabricated several optics for the instrument but died before its completion; it is a joint project of the Hobby-Eberly Telescope partnership and the Instituto de Astronom\'{\i}a de la Universidad Nacional Aut\'onoma de M\'exico.

This research has made use of the NASA/IPAC Extragalactic Database (NED) which is operated by the Jet Propulsion Laboratory, California Institute of Technology, under contract with the National Aeronautics and Space Administration.

\bibliographystyle{apj}
\bibliography{all.010115}

\appendix
\section{Evaluation of Uncertainties in Spectral Measurements}
\label{app:measurements}
In this appendix we carry out a variety of tests to quantify the uncertainties in the quantities reported in this paper, including the moments of the line profiles and other properties of the broad emission lines. We pay particular attention to uncertainties arising from the decomposition of the observed spectra since the method one uses for the decomposition may lead to systematic errors in the results  \citep[e.g.,][]{denney11,runnoe13a}. We begin by comparing the results from two different decomposition methods, the one adopted in this paper and the one adopted in \pone. We then consider how the measurement of moments of line profiles is influenced by noise in the wings of the profiles, which leads us to define an optimal spectral window around the peak of the line profile for making such measurements. Finally we examine the differences resulting from making measurements on the observed spectra and on the parametric models that describe these spectra and close with a comparison of the profiles moments measured in \pone\ and those measured here.

\begin{figure*}[t]
\centerline{
\includegraphics[width=8.cm]{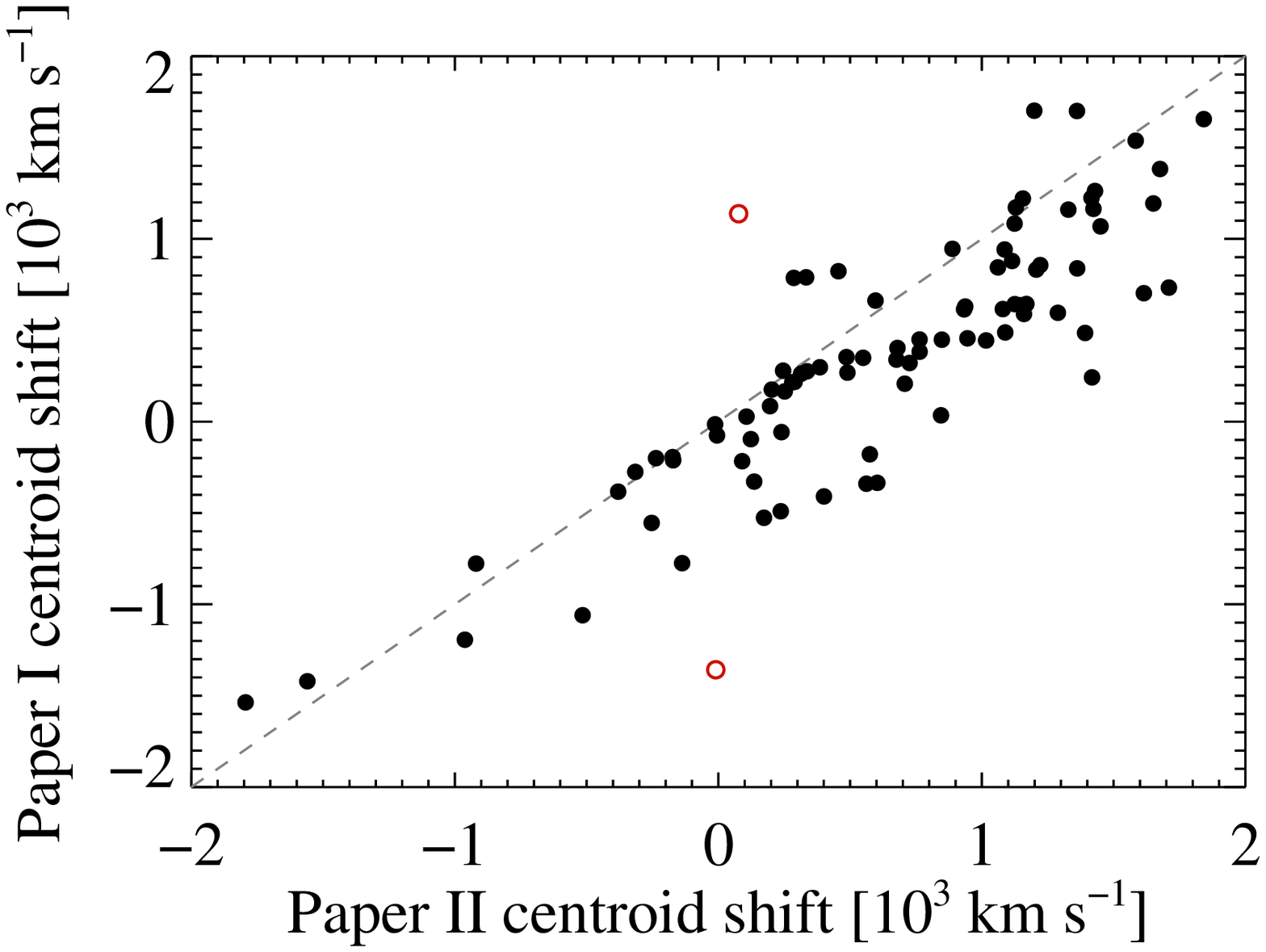}
\includegraphics[width=8.cm]{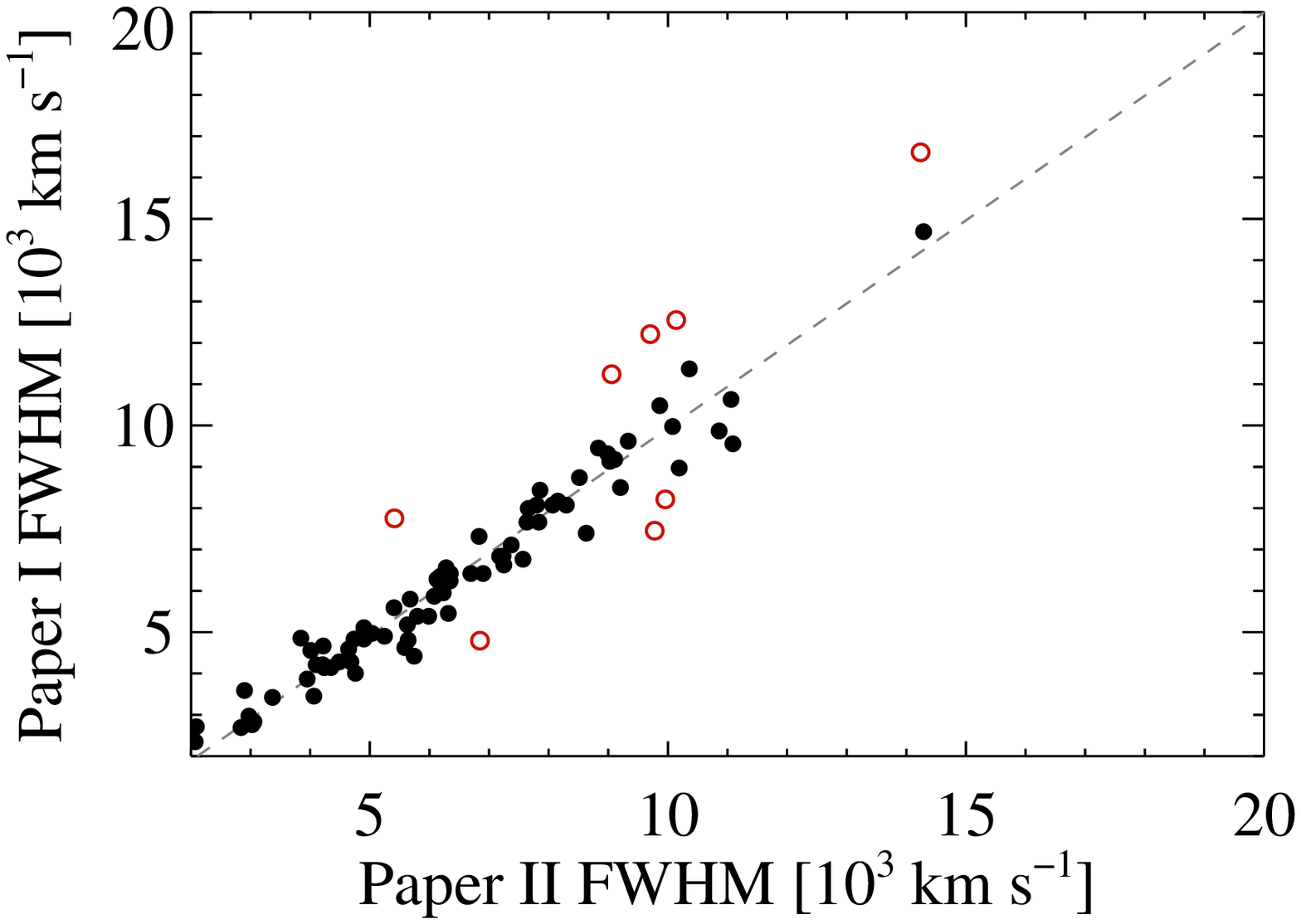}
}
\vskip -1cm
\centerline{
\includegraphics[width=8.cm]{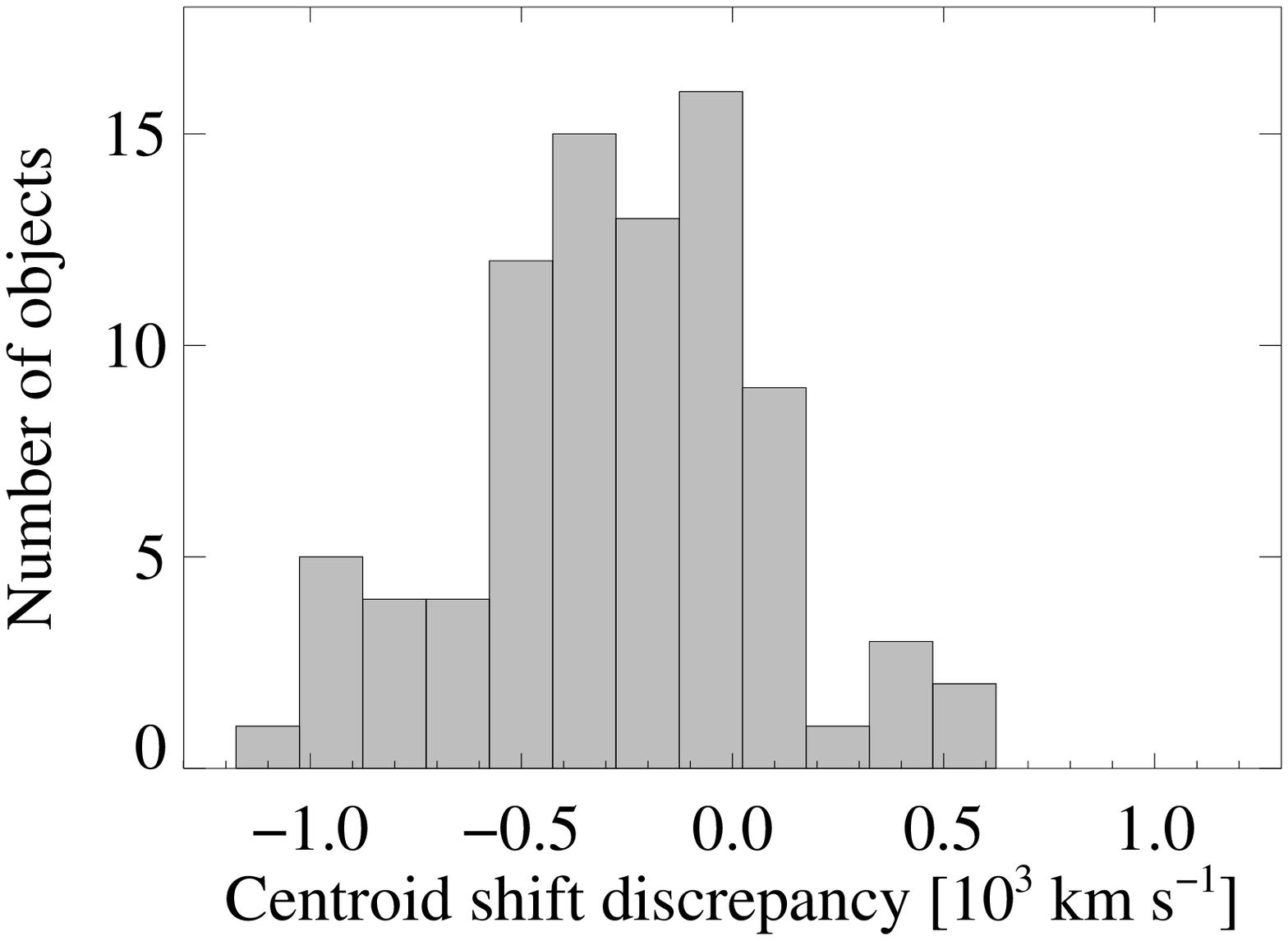}
\includegraphics[width=8.cm]{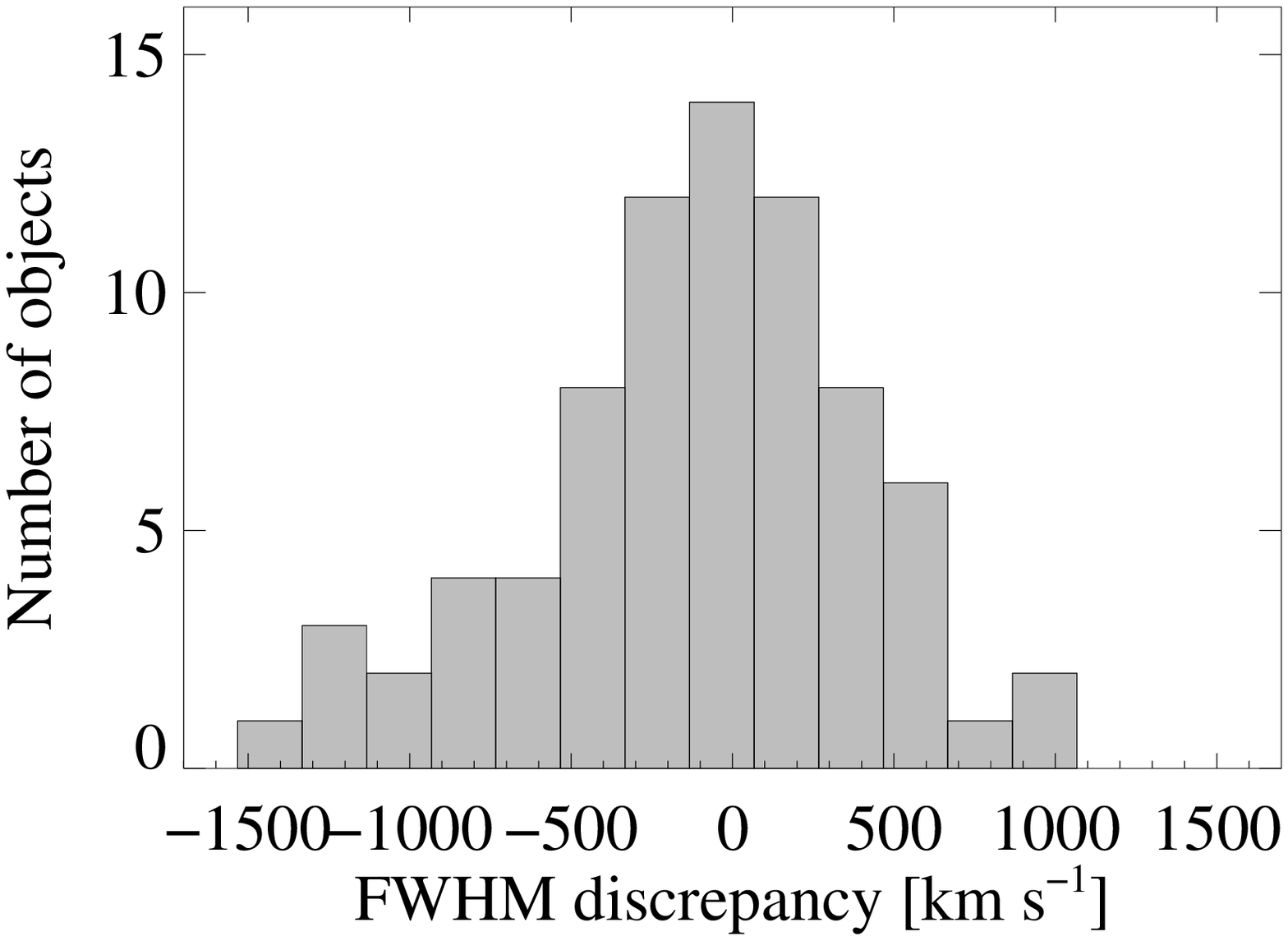}
}
\vskip -1cm
\centerline{
\includegraphics[width=8.cm]{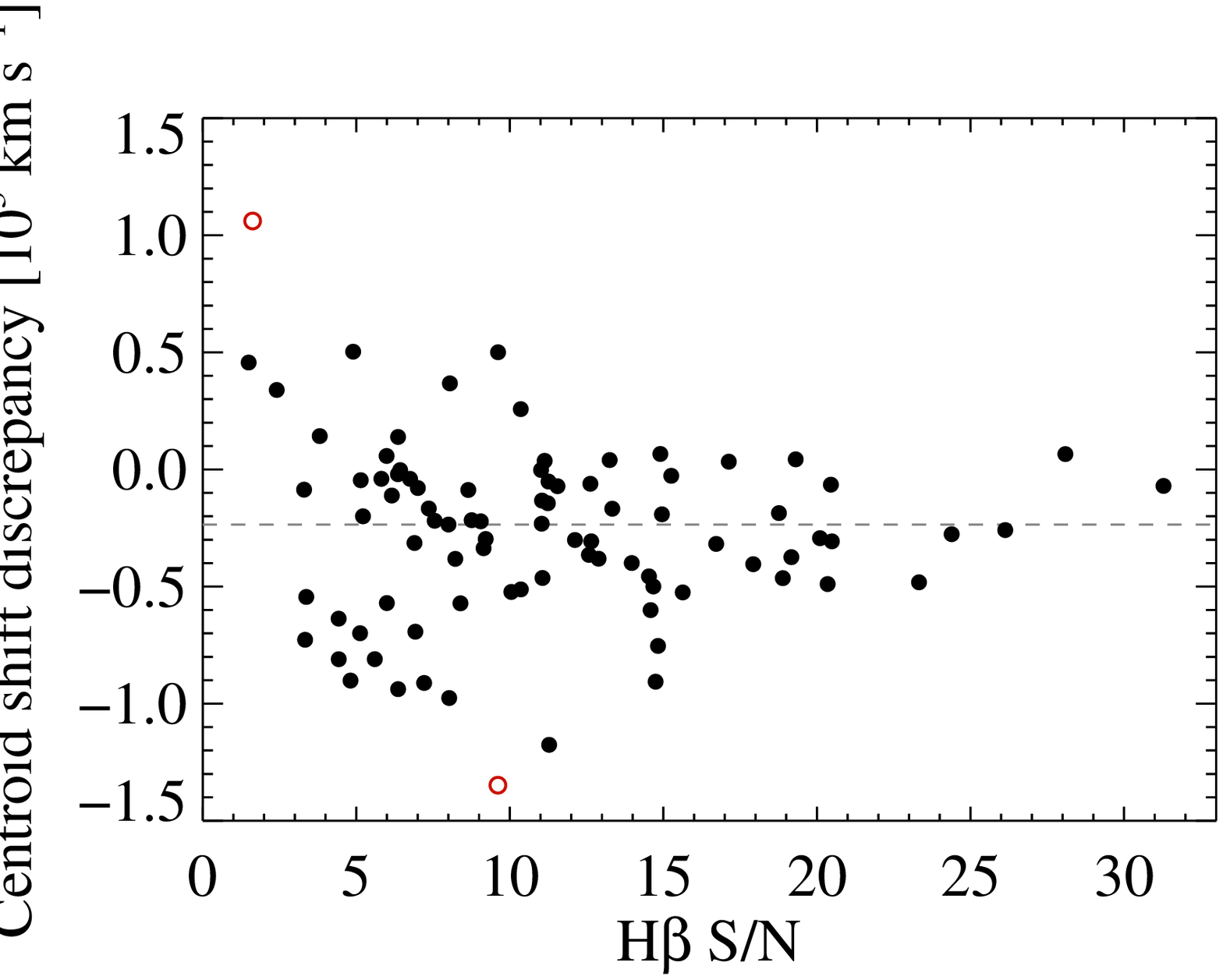}
\includegraphics[width=8.cm]{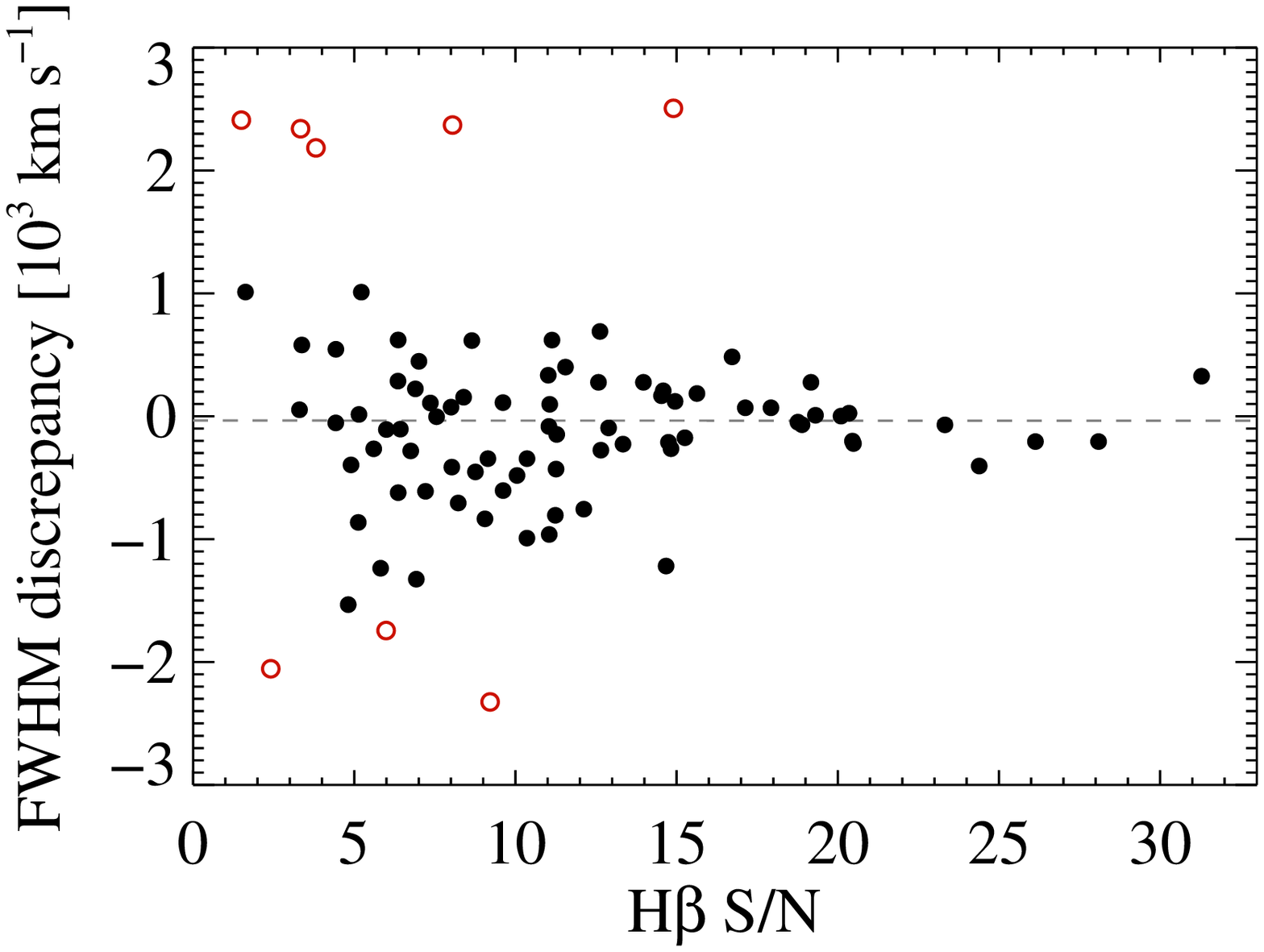}
}
\caption{A comparison of two spectral properties measured in identical fashion on broad \Hb\ spectra that were decomposed following the two different methods described in Appendix~\ref{app:dcomp}.  Top: A direct comparison of the measured properties.  The dashed line shows where the measurements are equal and red open points are rejected from statistics in Table~\ref{tab:dcomp_stats} by up to 5 iterations of $3\sigma$ clipping.   Middle: distribution of the measurement discrepancy described in Appendix~\ref{app:dcomp}.  Bottom: The discrepancy as a function of S/N in broad \Hb.  The dashed line shows the median value for the discrepancy after clipping.} \label{fig:dcomp} 
\end{figure*}

\subsection{The Effect of the Spectral Decomposition}
\label{app:dcomp}
\citet{denney09a} considered this issue for two objects observed multiple times during reverberation mapping campaigns and concluded in those cases that the measurements yield similar results. We are interested in this question generally, but also more specifically for objects with shifted broad emission lines.

We use the SDSS spectra from our sample to address this question because we have decomposed them twice, first following the method described in \pone\ and again following the new method described in Section~\ref{sec:sfit}.  There are two significant differences between these methods.  Firstly, \pone\ uses the \citet{bg92} \FeII\ template whereas here we adopt the template of \citet{veron-cetty04}.  The primary difference between these templates is that \citet{veron-cetty04} identified a region of narrow lines and excluded these from their \FeII\ template.  In practice, the two templates are similar with a small difference around 5000~\AA\ that can be important, particularly when the \FeII\ emission lines are  strong and/or narrow.  Secondly, we adopt a different procedure for subtracting the narrow lines than in \pone.  In both cases, an \OIII\ line is used to constrain the shape of the narrow \Hb.  In \pone, a narrow line template was constructed by fitting the \OIII\ profiles with a cubic spline after fitting the underlying wing of the broad \Hb\ line with a low-order polynomial.  Here we fit the  \OIII\ profiles with combination of Gaussians instead. In practice, the method adopted in \pone\ tends to subtract more of the underlying flux in the \OIII\ region than the one adopted here because no assumption is made about the shape the \OIII\ wings.

In order to make measurements of the spectral quantities of interest (line widths, shifts, fluxes, and profile models; see Table~\ref{tab:dcomp_stats}), we isolate the broad \Hb\ profile by subtracting the continuum, and the \FeII, \OIII\ and narrow \Hb\ lines. The measurements are made directly on the data following the procedure adopted in \pone\ and also on the best-fitting model for the broad \Hb\ profile as in the procedure adopted in this paper. The resulting measurements for two representative quantities are compared Figure~\ref{fig:dcomp}.  In this figure, the top row shows the measurements from each decomposition method plotted against each other with a unit slope line superposed for reference. The middle row shows the distribution of discrepancies between the measurements.  We adopt a convention where the discrepancy in a quantity $X$ is always taken to be $\Delta X = X_{\rm{\pone}}-X_{\rm{\ptwo}}$ and the fractional discrepancy is $\Delta X/\langle X\rangle$. The bottom row shows the relation between the discrepancy between measurements and the $S/N$ at the peak of the broad \Hb\ line. Statistical measures describing these distributions are summarized in Table~\ref{tab:dcomp_stats}.

We find that the distributions of discrepancies between measurements often have a small number of significant outliers, which we reject before calculating statistics for the distributions. We iteratively reject $>3\sigma$ outliers relative to the median until one of the following three conditions is met: a) there are no $>3\sigma$ outliers left b) fewer than 1\% of the total points have been rejected in the last iteration, or c) we have repeated the process five times. The rejected points are shown as red open circles in the relevant scatter plots and are included in the relevant histograms.  These points are excluded when calculating the statistics reported in Table~\ref{tab:dcomp_stats} and their fractional number is included in the last column of the table.

To investigate whether the magnitude of the discrepancies depends on the $S/N$ at the peak of the broad \Hb\ line, we evaluate the root-mean-square dispersion about the best-fitting parametric model near the peak of the line and plot it against the discrepancy in the bottom row of Figure~\ref{fig:dcomp}. In most cases displayed in these figures, the discrepancies between methods are smallest at the highest $S/N$ and become larger at lower $S/N$.

The measurements based on different spectral decompositions agree reasonably well.  Trends in the measurement discrepancies (i.e., asymmetries in the histograms of Figure~\ref{fig:dcomp}) can be attributed to the difference in the \OIII\ subtraction.  The method of \pone\ depresses the wing of the broad \Hb\ line more than the method of \ptwo\ and makes the profiles cuspier. Thus the measurements from \pone\ give bluer centroid wavelengths, somewhat larger skewness coefficients, and smaller FWQM for the broad \Hb\ lines. Moreover, the \OIII\ lines in \pone\ have somewhat larger EWs. The relative velocities measured by cross-correlating broad \Hb\ spectra are not affected, although the absolute velocity of the broad \Hb\ lines measured from the original SDSS spectra incurs an uncertainty of $300\;{\rm km\;s}^{-1}$ as a result of the different methods of fitting the profiles around the peak in Papers~\textsc{I} and \textsc{II}.

\begin{deluxetable}{lrrrcc}[t]
\tablecolumns{10}
\tablewidth{0pc}
\tablecaption{Measurement discrepancies between spectral decomposition methods\,\tablenotemark{a}
\label{tab:dcomp_stats}
}
\tablehead{
\colhead{} & 
\colhead{} & 
\colhead{} & 
\colhead{Standard} & 
\colhead{Fractional} & 
\colhead{Outlier} \\ 
\colhead{Property} & 
\colhead{Mean} & 
\colhead{Median} & 
\colhead{Deviation} & 
\colhead{Difference\,\tablenotemark{b}} & 
\colhead{Fraction\,\tablenotemark{c}} \\ 
\colhead{(1)} & 
\colhead{(2)} & 
\colhead{(3)} & 
\colhead{(4)} & 
\colhead{(5)} & 
\colhead{(6)} 
}
\startdata
\hline
Centroid Velocity Shift (km~s$^{-1}$)
&
$-$300
&
$-$200
&
300
& \nodata &
0.02
\\
Velocity Dispersion (km~s$^{-1}$)
&
5
&
50
&
200
&
0.08
&
0.05
\\
Skewness Coefficient
&
$-$0.08
&
$-$0.04
&
0.18
& \nodata &
0.10
\\
Pearson Skewness Coefficient
&
$-$0.02
&
$-$0.01
&
0.03
& \nodata &
0.10
\\
Kurtosis Coefficient
&
0.10
&
0.06
&
0.20
&
0.07
&
0.11
\\
FWHM (km~s$^{-1}$)
&
$-$100
&
$-$70
&
500
&
0.09
&
0.09
\\
FWQM (km~s$^{-1}$)
&
$-$600
&
$-$500
&
1000
&
0.13
&
0.02
\\
Peak Velocity Shift (km~s$^{-1}$)
&
$-$30
&
$-$20
&
300
& \nodata &
0.06
\\
Int. flux (erg~s$^{-1}$~cm$^{-2}$)
&
$-$60
&
$-$40
&
100
&
0.11
&
0.07
\\
EW (\AA)
&
9
&
8
&
10
& \nodata &
0.03
\\
\FeII\ int. flux (erg~s$^{-1}$~cm$^{-2}$)
&
70
&
70
&
200
& \nodata &
0.05
\\
\FeII\ EW (\AA)
&
7
&
6
&
11
& \nodata &
0.02
\\
\OIII\ int. flux (erg~s$^{-1}$~cm$^{-2}$)
&
$-$4
&
$-$1
&
40
& \nodata &
0.06
\\
\OIII\ EW (\AA)
&
5
&
3
&
7
& \nodata &
0.07
\\
$f_{\lambda}$(5100~\AA) (erg~s$^{-1}$~cm$^{-2}$ \AA$^{-1}$)
&
$-$2.9
&
$-$2.3
&
2.7
&
0.20
&
0.06
\\
\enddata
\tablenotetext{a}{Statistics after outlier rejection for the distributions of measurement discrepancies shown in Figure~\ref{fig:dcomp}.  The discrepancy in a quantity $X$ is taken to be $\Delta X = X_{\textrm{Pap\,{\sc i}}} - X_{\textrm{Pap\,{\sc ii}}}$.}
\tablenotetext{b}{The listed value is the standard deviation of the fractional difference distribution, where the fractional difference is $\Delta X/ \langle X \rangle$.}
\tablenotetext{c}{The outlier fraction is the fraction of objects rejected after up to 5 iterations of 3$\sigma$ clipping.}
\end{deluxetable}


\subsection{Practices for Measuring Spectral Properties}
\label{app:bwin}
The practices adopted for measuring spectral properties can vary between authors. Most notably, some authors make the measurements directly from the observed spectra while others fit parametric models to the spectra and make the measurements from the models (e.g., \citealt{vestergaard06} versus \citealt{shang07}). Another variation in practices is the wavelength around a broad emission line profile used to make the measurements. We determine the best wavelength window for making spectral measurements by systematically evaluating measured spectral properties for two windowing methods: a fixed window, where a hard wavelength limit is adopted for all objects, and a flexible window that extends from peak wavelength until the flux has dropped to some percentage of the line peak.  For the model spectra, we find that the best window is, $4600-5100$~\AA\ for a fixed-width scheme and down to 1\% of the peak flux for a flexible scheme.  We prefer the window based on a percentage of the peak flux because of its flexibility; the window can expand to accommodate very broad lines, asymmetric lines, or, in the case of this sample, very shifted lines.  In the case of measurements made on the data, a more ``conservative'' range is required for the flexible window so that the measurement is not influenced by noise.  When making measurements on the observed data, it is therefore best to use a window going out from the line center until the spectrum reaches 7\% of the peak flux.  Measurements involving the height the line peak (e.g., FWHM) are not particularly sensitive to the window, whereas the moments of the line profile and integrated flux do depend on the window; the consequences of using too wide a window are much more significant in measurements made on the data.  Following the methodology adopted for comparing measurements from different spectral decompositions, we investigate the differences between making measurements on the data versus the best-fitting model.  In general, though it appears that measurements made on the data are more sensitive to the characteristics of and spikes in the noise, we find that the overall effect is negligible (e.g., the fractional difference in the velocity dispersion is 2\%) compared to the differences that can result from different spectral decomposition methods.

\subsection{A Comparison of Measurements in This Sample}
\label{app:comp}
\begin{figure}[t]
\centerline{
\includegraphics[width=8cm]{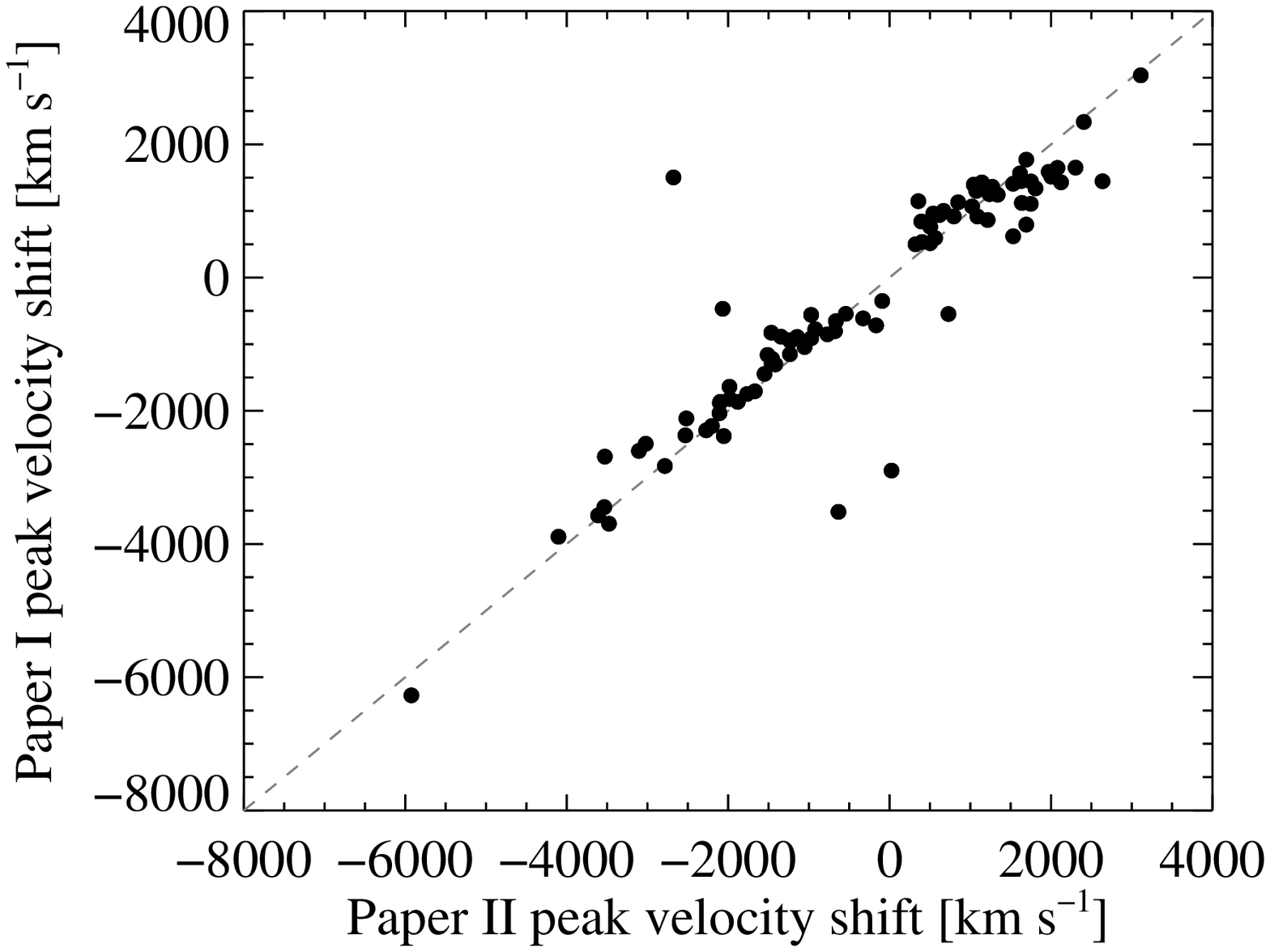}
\includegraphics[width=8cm]{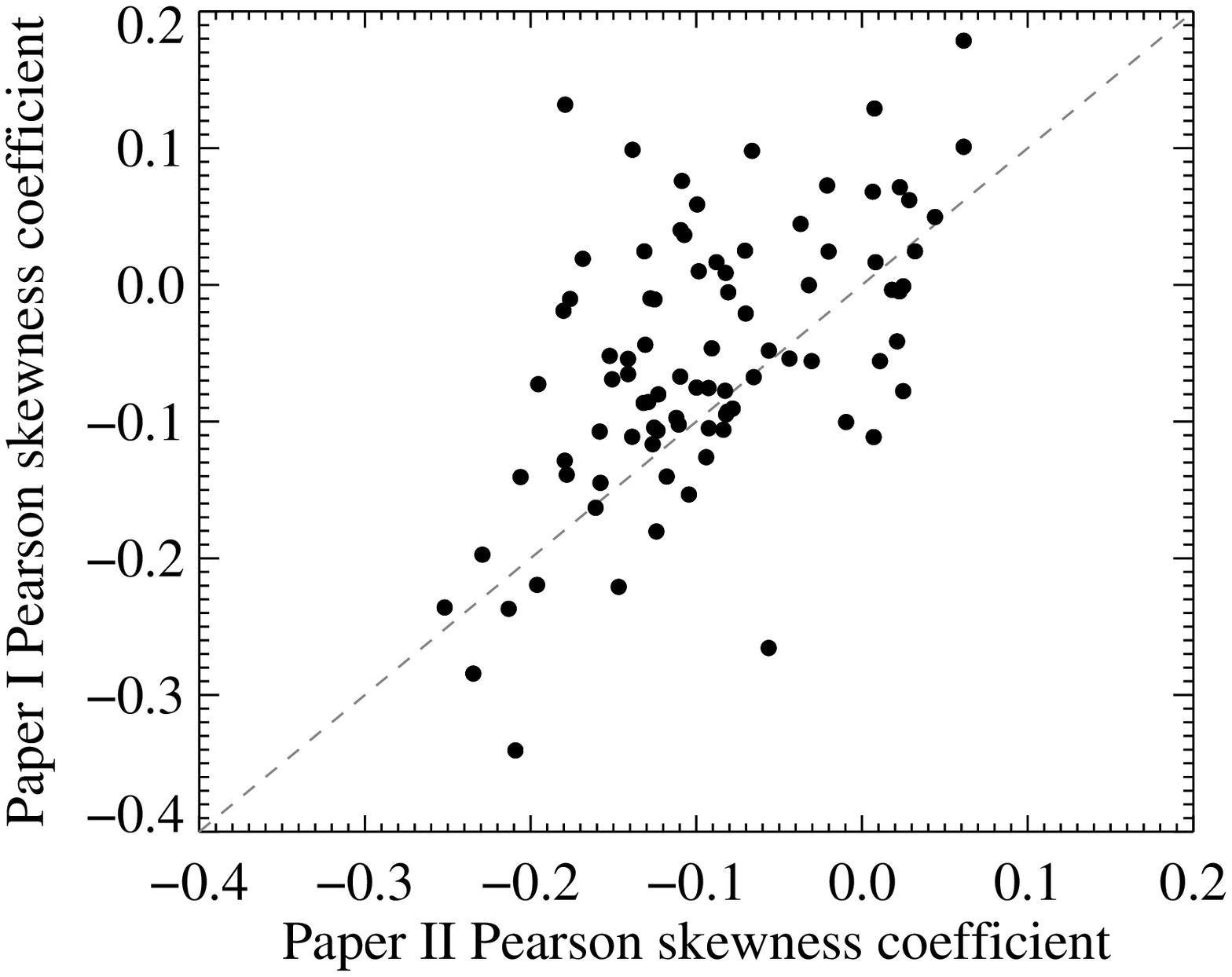}
}
\caption{A comparison of the velocity shift of the peak and Pearson skewness coefficient measured for the SDSS spectra in this work and in \pone.} \label{fig:jvm}
\end{figure}

\begin{figure}[h]
\epsscale{0.5} \plotone{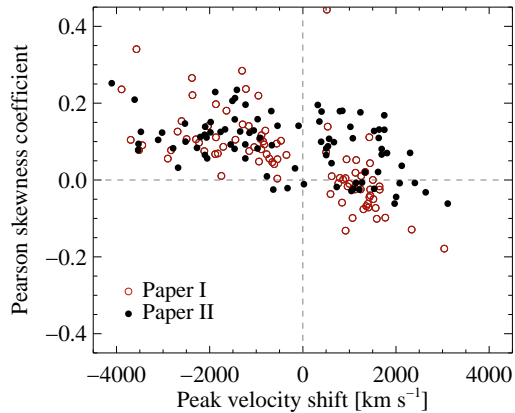} 
\caption{The correlation between Pearson skewness coefficient and the velocity shift of the peak found in \pone.  Open red circles show the measurements from \pone\ and the solid black circles show the re-measured values from this work.} \label{fig:pskewcorr}
\end{figure}

In this work we measure spectral properties on the best-fitting models from the decomposition described in Section~\ref{sec:sfit} in the wavelength window where the model is greater than 1\% of the peak model flux.  This differs from \pone\ in the spectral decomposition, the wavelength window of interest, and the spectrum on which the measurements are made.  As a result, there is scatter between the properties originally measured for the SDSS spectra and those we adopt here.  Measured spectral properties are listed here in Tables~\ref{tab:hbmeasurements} and \ref{tab:o3measurements} and can be compared to the original values in tables 3 and 4, respectively, in \pone.  For the velocity shift of the peak and the Pearson skewness coefficient, we compare the \pone\ measurements with our adopted values in Figure~\ref{fig:jvm}.  These properties are of particular interest because the velocity shift of the peak anchors the radial velocity curves and because of the correlation between the Pearson skewness coefficient and velocity shift of the peak identified in \pone.  The correlation shown in Figure~5 of \pone\ illustrates the tendency for profiles to be asymmetric in the direction of their shift.  We find that this correlation is robust against the details of the spectral decomposition and measurement and show it here in Figure~\ref{fig:pskewcorr}.  For the measurements in this work, the Spearman Rank correlation coefficient is $\rho=0.419$, with a probability $P=7.7\times10^{-5}$ that this distribution of points would be found by chance.  Thus, while the details of the correlation do depend on the measurements, the general trend is robust and statistically significant.



\label{lastpage}
\end{document}